\documentclass[lettersize,journal]{IEEEtran}
\usepackage{amsmath,amsfonts}
\usepackage{algorithm}
\usepackage{array}
\usepackage[caption=false,font=normalsize,labelfont=sf,textfont=sf]{subfig}
\usepackage{textcomp}
\usepackage{stfloats}
\usepackage{url}
\usepackage{verbatim}
\usepackage{graphicx}
\usepackage{cite}
\usepackage{multirow}
\usepackage{algpseudocode}
\usepackage{hyperref}
\hypersetup{
    colorlinks=true,
    linkcolor=blue,
    filecolor=magenta,      
    urlcolor=cyan,
    }
\usepackage{soul,color}
\usepackage[normalem]{ulem}

\begin{document}

\title{What, When, Where to Compute-in-Memory for Efficient Matrix Multiplication during Machine Learning Inference}

\author{Tanvi Sharma, Mustafa Ali, Indranil Chakraborty, and Kaushik Roy, \IEEEmembership{Fellow,~IEEE,}
\thanks{This work has been submitted to the IEEE for possible publication. Copyright may be transferred without notice, after which this version may no longer be accessible.}
}

\markboth{}%
{Shell \MakeLowercase{\textit{et al.}}: A Sample Article Using IEEEtran.cls for IEEE Journals}


\maketitle

\begin{abstract}

  Matrix multiplication is the dominant computation during Machine Learning (ML) inference. To efficiently perform such multiplication operations, Compute-in-memory (CiM) paradigms have emerged as a highly energy efficient solution. However, integrating compute in memory poses key questions, such as  1) \textit{What type of CiM to use}: Given a multitude of CiM design characteristics,
    determining their suitability from architecture perspective is needed. 2) \textit{When to use CiM}: ML inference includes workloads with a variety of memory and compute requirements, making it difficult to identify when CiM is more beneficial than standard processing cores. 3) \textit{Where to integrate CiM}: Each memory level has different bandwidth and capacity, creating different data reuse opportunities for CiM integration. 
    
    To answer such questions regarding on-chip CiM integration for accelerating ML workloads, we use an analytical architecture-evaluation methodology with tailored mapping algorithm. The mapping algorithm aims to achieve highest weight reuse and reduced data movements for a given CiM prototype and workload. Our analysis considers the integration of CiM prototypes into the cache levels of a tensor-core-like architecture, and shows that CiM integrated memory improves energy efficiency by up to $3.4 \times$ and throughput by up to $15.6 \times$ compared to established baseline with INT-8 precision. We believe the proposed work provides insights into \textit{what} type of CiM to use, and \textit{when} and \textit{where} to optimally integrate it in the cache hierarchy for efficient matrix multiplication.

\end{abstract}


\section{Introduction}

\begin{figure}[!t]
  \centering
  \includegraphics[ width=\columnwidth]{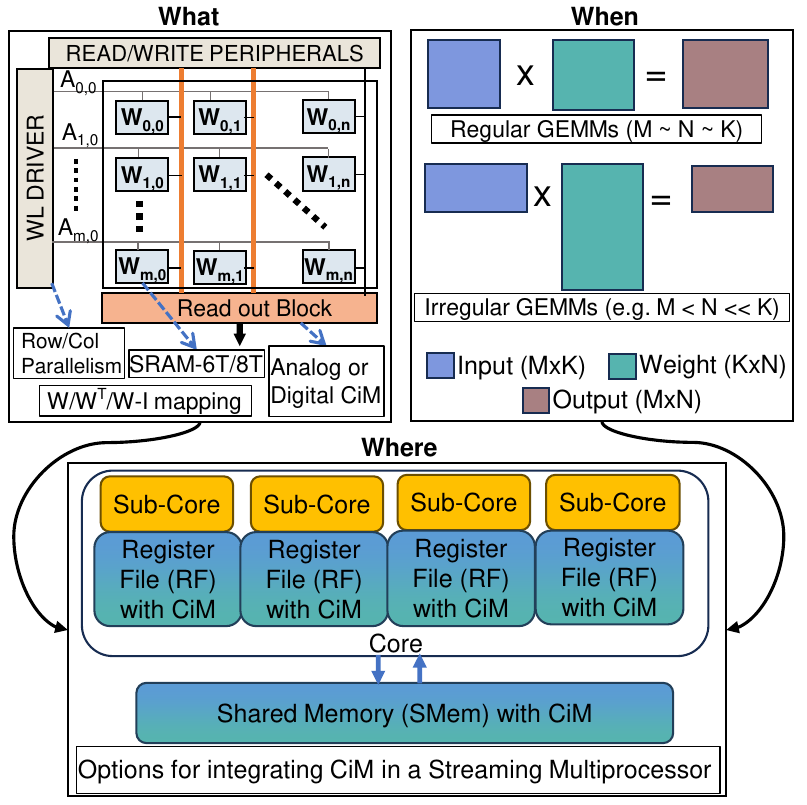} 
  \caption{Overview figure showing the challenges that need to be addressed when computing in on-chip memory. There is a need to understand the difference in characteristics of various CiM types (What), GEMM shapes (When) and memory levels (Where) for optimal ML inference acceleration.
  }
  \label{fig:intro}

\end{figure}


Machine learning (ML) applications have become ubiquitous across various domains such as automotive, health care, finance, and technology.
This has led to an increase in demand for high-performance and energy-efficient ML hardware solutions. 
 Most hardware solutions focus on efficiently executing the core computations involved in ML workloads, specifically the matrix-vector multiplications and general matrix-matrix multiplications, also known as GEMMs\cite{welser2018future, kim2023full}. 
 However, due to the separation of processing units from memory in von Neumann architectures such as Central Processing Units (CPUs) and Graphics Processing Units (GPUs), such data intensive multiplications incur high energy costs due to frequent memory accesses. The resulting large data movement, between the processing unit and memory, is commonly known as the ``memory wall" or ``von-Neumann bottleneck"
 ~\cite{wulf1995hitting}. 
To alleviate such bottleneck, Compute-in-Memory (CiM) paradigms have been proposed to reduce the expensive data movement costs and provide energy-efficient solutions by performing computations in the memory \cite{verma2019memory}.

There are numerous ways of integrating CiM across the memory hierarchy: from CMOS on-chip cache memory to DRAM or Flash \cite{seo2023advances, mutlu2022modern, 9435945}.  
In this work, we focus on seamless integration of CiM in on-chip memory subsystem. While integrating CiM in caches has been explored \cite{fujiki2019duality, fujiki2022multi,lockerman2020livia}, a comprehensive evaluation of the effectiveness of different types of CiM designs at the architecture level, particularly for ML inference, is yet to be studied. Our work explores the benefits of integrating CiM to different cache levels, Register File (RF) and Shared Memory (SMEM), in a Streaming Multiprocessor (SM) of a GPU \cite{ampere}. 
To efficiently utilize CiM in a memory subsystem, there is a need to determine the optimum type of CiM, when to use it, and where to use it, for ML inference (Fig. \ref{fig:intro}). Note, we use the term `primitive' to refer to a given CiM design, inspired from the use of `primitives' for linear algebra functions \cite{buluc2010linear}. Next, we discuss the challenges in determining the answers to the what-when-where questions.

%

\textbf{Challenge 1 (What type of CiM):} The increase in CiM related research over the past few years has lead to a variety of techniques to enable multiply and accumulate (MAC) operation in memory\cite{computesram,analog1cim,ali2023cicc,digitaltsmc,tmsc2023digital,isscc2022mfchang,tsmc2022isscc}. Broadly, CiM can be classified as analog or digital, based on their nature of computing. Analog CiM performs MAC operations in the analog/mixed signal domain inside a memory array, by accumulating current or charge. In contrast, digital CiM performs all the computations in the digital domain by including bit-wise logic gates such as AND or XOR near the sensing circuitry. On one hand, analog CiM requires heavy circuit blocks such as Digital-to-Analog (DAC) and Analog-to-Digital (ADC) converters for robust communication among different CiM blocks. ADCs increase the overhead of analog CiM due to their high area and latency costs. Digital CiM, on the other hand, would have a higher compute latency due to multiple bit-wise operations and accumulations to calculate the full precision MAC output. Also, design choices such as the type of memory cell (SRAM-6T/8T), the number of wordlines or bitlines enabled at a time, and the mapping scheme for weights in the memory array have made it increasingly challenging to identify the most effective CiM primitive in a system (Fig. \ref{fig:intro}).


\begin{table*}[t]
  \centering
 \caption{Description of GEMMs(M,N,K) found during ML Inference$^{*}$}
        
  \label{table:mlworkloads}
  \small
    \begin{tabular}{|l|l|l|l|}
    \hline
    ML Workload & M & N & K \\
    \hline
    Convolution2D Layer & $Height_{o}$$\times$$Width_{o}$ & $Channel_{o}$ & $Height_{i}$$\times$$Width_{i}$$\times$$Channel_{i}$ \\
    Fully Connected Layer    & $Output~Dimension$     & $Batch$ & $Input~Dimension$ \\
    Attention Layer-K\//V & $Embedding~Size$ & $Seq~Length$ & $Embedding~Size$ \\
    Attention Layer-QK\textsuperscript{T} & $Seq~Length$ & $Seq~Length$ & $Embedding~Size$ \\
    Attention Layer-QK\textsuperscript{T}V & $Embedding~Size$ & $Seq~Length$ & $Seq~Length$ \\
    \hline
    \multicolumn{4}{l}{$^{*}$Assuming single batch size and fused matrix-multiplication during attention score calculation.}
    \end{tabular}%

\end{table*}

\begin{figure}[!t]
  \centering
  \includegraphics[width=0.43\textwidth]{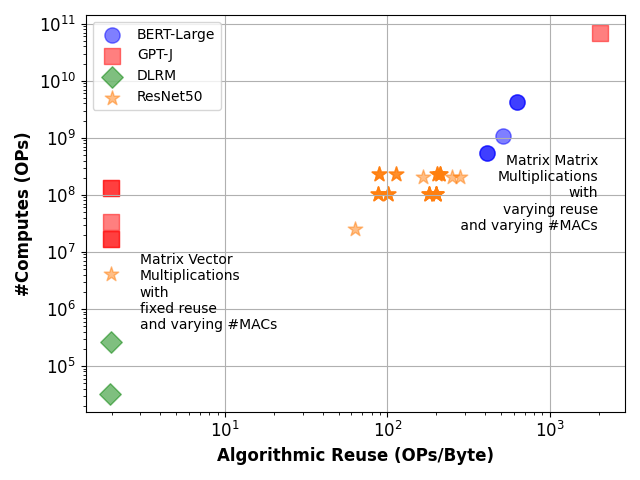} 
  \caption{Comparison of different GEMM shapes in terms of operations ($2 \times M \times N \times K$)  and algorithmic reuse (operations\//byte) to show the memory-intensive and compute-intensive nature of matrix multiplications during ML inference, assuming INT8 precision and single batch size. The darker shades of scatter points imply higher frequency of occurrence of the same GEMM shape for that workload. }
  \label{fig:gemms}
    
\end{figure}

\textbf{Challenge 2 (When to use CiM)}: 
ML inference is dominated by matrix-vector multiplications and general matrix-matrix multiplications, collectively referred as GEMMs in this paper. GEMM ($M\times N \times K$) computation can be thought of as multiplying input matrix of size $M \times K$ with weight matrix of size $K \times N$ to get an output matrix of size $M \times N$ \cite{chetlur2014cudnn} (Fig. \ref{fig:intro}). Table \ref{table:mlworkloads} describes the shapes of GEMMs for different ML workload layers. Further, Fig. \ref{fig:gemms} shows the compute and memory requirements of different types of GEMMs present in convolution, transformer and recommendation layers. Here, \textit{algorithmic reuse} is the ratio of arithmetic operations to memory accesses, and can be calculated as the number of MAC operations divided by the total size of matrices. Algorithmic reuse highlights the maximum reuse opportunities for each GEMM. However, to achieve highest possible reuse, it is important to efficiently \textit{map each} GEMM to the CiM primitives, and decide when the benefits of CiM are higher than GPU cores.

%
\textbf{Challenge 3 (Where to integrate CiM)}:
Since GEMMs have regular data access patterns, they offer high temporal and spatial locality. Typically, GPUs optimize their memory hierarchy to efficiently leverage such locality by fetching the matrices from main memory in blocks or smaller tiles\cite{gemmnv} (Fig. \ref{fig:tiling}). In addition, GPUs parallelize such tiled GEMM operations across hundreds of tensor cores, present in sub-cores of SMs. 
CiM based hardware designs are also capable of performing parallel matrix multiplications, by enabling multiple columns and rows inside multiple memory arrays. However, each memory level differs in terms of bandwidth and storage capacity, which affects the data reuse opportunities and compute parallelism when integrating with CiM features. Hence, it is crucial to find the memory level that can effectively exploit the locality and provide highest benefits for CiM integration.

\textbf{Our approach:}
To address the above challenges, we develop a systematic methodology to perform a fair comparison across various CiM primitives, workloads and memory levels. First, to accommodate different CiM characteristics for analytical evaluation, we devise a dataflow-centric representation of CiM primitives, by breaking down the design into smaller CiM units. This helps in easily abstracting out CiM specifications in terms of compute parallelism and compute energy for measuring performance and energy efficiency at the system level. Further, to tackle the challenge of mapping GEMMs to CiM based architectures, we propose a dataflow algorithm that aims to maximise the weight stationary benefits of CiM and reducing data movement. Subsequently, we perform a comprehensive evaluation, assuming iso-area constraints, with typical CiM primitive and different cache levels for various shapes of GEMM, based on both real and synthetic workloads. Such analysis assists in finding answers to the mentioned questions and developing strategies for optimal CiM based architecture designs. The main contributions of the work can be summarized as follows:


\begin{itemize}
    \item Analytical evaluation of SRAM-based CiM primitives at Register File (RF) and Shared Memory (SMEM) levels in a tensor-core-like processor architecture.

    \item Dataflow algorithm to optimize performance and energy efficiency gains through optimal mapping 
    for a given CiM architecture and GEMM shape.
    
    \item Detailed analyses to the questions on what, when and where to CiM for various GEMM shapes from energy/performance perspective.
    
\end{itemize}
 The rest of the paper is organized as follows: Section \ref{sec:related} distinguishes our work from other studies done in the past. Section \ref{sec:background} provides relevant background for this work. Section \ref{sec:main} describes the dataflow-centric approach used for representing CiM enabled caches and our priority based mapping algorithm. Section \ref{sec:methodology} describes the experimental setup, followed by 
Section \ref{sec:results}, which highlights the results and key takeaways on designing SRAM-based CiM architectures. The last section concludes the work by providing a short summary. 

\section{Related Works}
\label{sec:related}

While in-cache computing has been explored in CPUs \cite{fujiki2019duality}, integrating compute into the GPU memory subsystem for ML inference remains unstudied. Duality Cache \cite{fujiki2019duality} accelerates data-parallel applications in the last-level cache of Xeon processors, and MLIMP \cite{fujiki2022multi} extends this for graph neural networks with a concurrent task scheduler for multi-layer in-memory processing.
On the other hand, this work focuses on analyzing the benefits of integrating compute at different levels of a GPU memory hierarchy for ML inference. We consider GPUs because of their dominance in accelerating GEMMs, the core computation in inference tasks. In addition, GPUs are programmable accelerators and the same programming model could be potentially re-used for CiM integrated GPUs.

Livia \cite{lockerman2020livia} also modified cache memory in CPUs to minimize data movement for irregular-data-access applications. They propose a system architecture that dynamically schedules tasks and data at different locations in the memory hierarchy. In contrast, we focus on both regular and irregular workloads (GEMMs) and compare the benefits of CiM parallelism with standard architectures.

To-Pim-or-Not \cite{devic2022pim} is the first work to raise the questions on how, and when to use processing-in-memory (PIM) for different applications, focusing on the development of a software framework for the same.
However, its scope is limited to emerging general purpose DDR Memory Systems rather than SRAMs, which is the focus of our work. A recent work \cite{houshmand2023benchmarking} on benchmarking analog vs. digital compute in memory develops a quantitative energy model based on fixed analog CiM and digital CiM designs, referred to as templates. 
However, the study lacks a comprehensive evaluation of how the workload shape affects the CiM benefits. 

\section{Background}
\label{sec:background}
\begin{figure}[t]
    \centering
    \includegraphics[ width=\columnwidth]{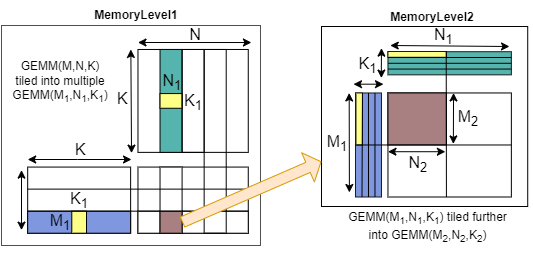}
    \caption{GEMMs are tiled (or blocked) into smaller tiles in the memory hierarchy to exploit spatial and temporal locality.}
    \label{fig:tiling}
\end{figure}
\subsection{Importance of GEMMs in ML Workloads}
Machine learning workloads consist of a broad range of neural networks from convolutional, fully connected, to transformer and recommendation models. During ML inference, input, in the form or text or image, is fed to such neural networks to compute the final predicted output. Hence, inference only consists of a single forward pass through the network. In contrast, ML training phase constitutes both forward pass and back-propagation to calculate the weight updates. Since training requires frequent weight updates and writing to memory, we only focus on inference tasks in this work.

Matrix-vector multiplication and matrix-matrix multiplications are at the core of computation in these neural networks \cite{welser2018future,kim2023full}. From now onwards, we refer to such multiplications under one umbrella term, known as general matrix-matrix multiplications or GEMMs ($M\times N \times K$). \textit{$M$ and $N$ represent the dimension of the output matrix and $K$ is the reduction or accumulation dimension} (Table \ref{table:mlworkloads}). For legacy purposes, we still use input, weight and output matrix to represent $M\times K $, $K \times N$ and $M \times N$ matrices, respectively, in the paper. 

Convolution neural networks (CNNs) can be implemented as GEMMs by transforming the convolution operation of input and weight feature maps to matrix-matrix multiplication using im2col \cite{chetlur2014cudnn}. im2col or image-to-column transformation converts the 3D convolution operation to a GEMM ($M,N,K$) such that $K$ represents the reduction dimension for the MAC operation between inputs and weights, $M$ represents the total number of such convolutions and $N$ is decided based on the number of output channels. The initial layers of a CNN generally have larger input feature maps compared to other layers, for larger datasets such as ImageNet. The last layer is a classifier, which is essentially a fully connected (FC) layer. It consists of matrix vector multiplications, which can be thought of as a special case of GEMM with $M = 1$. Similarly, transformer models perform computation of the Query ($W_Q$), Key ($W_K$), and Value ($W_V$) matrices from input embedding in the attention layer. Additionally, transformer models comprise of other GEMMs such as logit ($QK\textsuperscript{T}$), attention ($QK\textsuperscript{T}V$), and output ($W_O$) calculations, followed by FC layers. On the other hand, recommendation models incorporate multilayer perceptrons (MLPs), to predict items from a pool of dense features and user preferences \cite{naumov2019deep}, basically consisting of FC layers.

\subsection{Dataflow during GEMM execution}

\begin{figure}[t]
    \centering
    \includegraphics[width=\columnwidth]{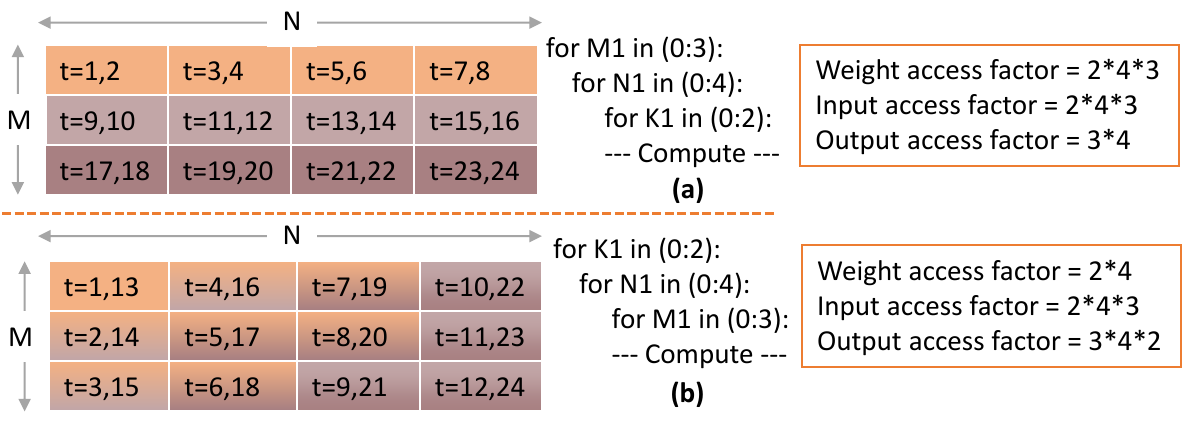}
    \caption{Two different dataflows with for-loop representation showing difference in the \textit{observed reuse}, decided by the number of data accesses. The time stamps are color coded to understand when the partial sums for the output matrix ($M$$\times$$N$) are calculated, creating opportunities for temporal reductions.}
    \label{fig:dataflow-effect}
\end{figure}


GEMMs exhibit high spatial and temporal locality owing to their regular data access patterns. Spatial locality refers to the use of data elements residing near each other, while temporal locality pertains to the repeated use of the same data elements over time. To exploit such locality, GPUs implement GEMMs by tiling (or blocking) the matrices as shown in Fig. \ref{fig:tiling}, and performing tile computations in parallel \cite{gemmnv}. The allocation and scheduling of these tiles is decided by the dataflow, represented using \textit{for}-loops. For a given dataflow, \textit{loop factor} explains the size of such tiles and \textit{loop order} (order of $M,N,K$ in loop representation of dataflow) decides the number of memory accesses or reuse of tile at a given memory level. Algorithmic reuse (arithmetic intensity) assumes that each matrix tile is fetched only once, and can be calculated as number of operations
divided by the total size of matrices fetched from the memory:
\begin{equation}
Algorithmic~reuse =\frac{2*(M*N*K)}{BP*(M*N + N*K + M*K)}    
\end{equation}
where BP is the bit-precision. 
However, \textit{observed} data reuse could be different than \textit{algorithmic} data reuse depending on the actual number of memory accesses. An example of how dataflow affects the observed reuse or, essentially, the number of memory accesses is shown in Fig. \ref{fig:dataflow-effect}.

The performance of GEMMs is limited by their algorithmic reuse and the hardware resources. Low arithmetic intensity GEMMs are limited by the memory bandwidth, while high arithmetic intensity GEMMs are limited by the peak performance. Libraries such as cuDNN and cuBLAS are used to decide the tile sizes for achieving highest possible performance for given GEMM shape. Larger tiles often run more efficiently on GPUs due to high reuse \cite{gemmnv}. This increase in data reuse can result in lower bandwidth requirements and improved efficiency. However, opting for larger tiles could reduce the number of tiles that can run in parallel due to limited hardware resources like shared memory or registers. This reduction can potentially lead to lower compute utilization and lower performance.

Given that the implementation of GEMMs is optimized on GPUs to get the best performance, it is important to achieve the best dataflow for CiM intergrated architecture as well. There are several studies for exploring the dataflow search space and choosing the optimal dataflow on ML accelerators. SCNN \cite{parashar2017scnn} was one of the first works which introduced dataflow optimization for deep neural networks (DNNs). They proposed an input-stationary dataflow, where input activation is held stationary, allowing it to be multiplied by all the filter weights necessary for each output channel. Further, Timeloop \cite{parashar2019timeloop} presented a low cost mapper and model to explore dataflow search space for DNNs and GEMMs. It models the input problem size as a nested loop, allowing for the assessment of data reuse opportunities and efficient mapping across different architectures and workloads. Maestro \cite{kwon2020maestro} is another tool that proposes an analytical cost model to assesses the cost-benefit tradeoffs in dataflow using their data centric approach. ZigZag \cite{mei2020zigzag} also explores the DNN accelerator design space by expanding the search to uneven scheduling opportunities. 


\subsection{SRAM based Compute-in-Memory Primitives}
Given the high cost of memory accesses compared to logic operations \cite{horowitz20141}, many works have been proposed to perform computations in on-chip SRAM \cite{shanbhag2022benchmarking}. These CiM macros can be designed in various ways 
-- analog or digital. Another key factor is the type of SRAM cells used. These cells vary in their transistor counts, with common examples including 6T \cite{isscc2022mfchang}, 8T \cite{ali2023cicc}, and 10T \cite{impulse} cells. Additionally, CiM macros vary in the way input data are stored or applied to CiM compute. For instance, input can be stored in the CiM macro itself or it can be applied to the CiM macro from an external buffer.

In digital CiM, the multiply and accumulate operations are performed in the digital domain through bit-serial logic gates. A sequence of such logic operations can be then combined to perform arithmetic operations. Such logic is usually placed in the peripheral circuitry of the CiM macro \cite{computesram,digitaltsmc,tmsc2023digital}. The degree of compute parallelism in digital CiM macros generally depends on the amount of logic resources added in the macro. However, adding more logic circuits to digital CiM designs leads to significant area overhead \cite{digitaltsmc}, resulting in performance/energy -- area trade-off. On the other hand, analog CiM macros perform MAC operations by applying input bits through wordlines while storing the weights in the CiM macro \cite{analog1cim,ali2023cicc}. The output is produced as an analog voltage or current at the bitlines, which needs to be converted to digital domain through an Analog to Digital Converter (ADC) for robust inter-macro communication. Notably, ADCs are the major area/latency/energy bottlenecks in analog CiM macros \cite{ankit2019puma}. Prior art tried to amortize the cost of ADCs for better energy-efficiency/performance through narrower output precision or novel ADC circuit designs \cite{analog1cim,ali2023cicc}. It is worth mentioning that digital CiM scales with the most advanced technology nodes \cite{tmsc2023digital} unlike analog CiM where ADCs suffer from significant noise at such advanced technology nodes \cite{tsmc7nmanalog}.


As mentioned before, CiM macros comprise various SRAM cell types. 8T-cells are commonly adopted in CiM macros since they have decoupled read and write ports, leading to minimal read-disturb issues \cite{computesram} and higher noise margin than 6T cells. 8T-based CiMs enable multiple wordlines simultaneously leading to more parallel MAC operations and energy efficiency. On the other hand, 6T-cells are the {\em de-facto} standard for conventional SRAM designs due their compact area. 6T-based CiM have been proposed to reduce the area overhead of 8T cells. To avoid the read disturb issue in 6T-based CiM macros, several circuit techniques have been proposed \cite{analog1cim, isscc2022mfchang, digitaltsmc}. For example, to perform 6T-based analog CiM, \cite{analog1cim,isscc2022mfchang} added a local computing block to a group of 6T cells that share the same bitline. There are multiple groups in a column where two cells from different groups do not share the same bitline. Note, only one 6T cell is activated per local computing group during computation to avoid read disturb. In addition to 6T and 8T based CiM, some reported macros adopted other cell types (e.g. 10T \cite{impulse}) that can perform more complex computations (such as in-memory addition and membrane potential updates for spiking neural networks) within the cell while leading to larger area overhead.

CiM macros also vary in the way inputs are applied. Input data can be stored in the CiM macro itself prior computations \cite{computesram}, or streamed in the macro during the CiM operation from an external buffer \cite{ali2023cicc}. Input storing/streaming imposes different mapping/dataflow constraints on the corresponding CiM macro leading to different optimal data orchestration.



Additionally, various input/output precisions have been shown in such works leading to comparisons challenges. To have a fair comparison, we fix input/output precision to 8-bit integer in this work. It is also worth mentioning that different CiM macros might impose certain dataflows at the macro level due to their unique compute nature \cite{computesram,analog1cim}. 

\section{Dataflow-Centric Approach for representing CiM Primitives}
\label{sec:main}

\begin{figure}[!t]
  \centering
  \includegraphics[width=\columnwidth]{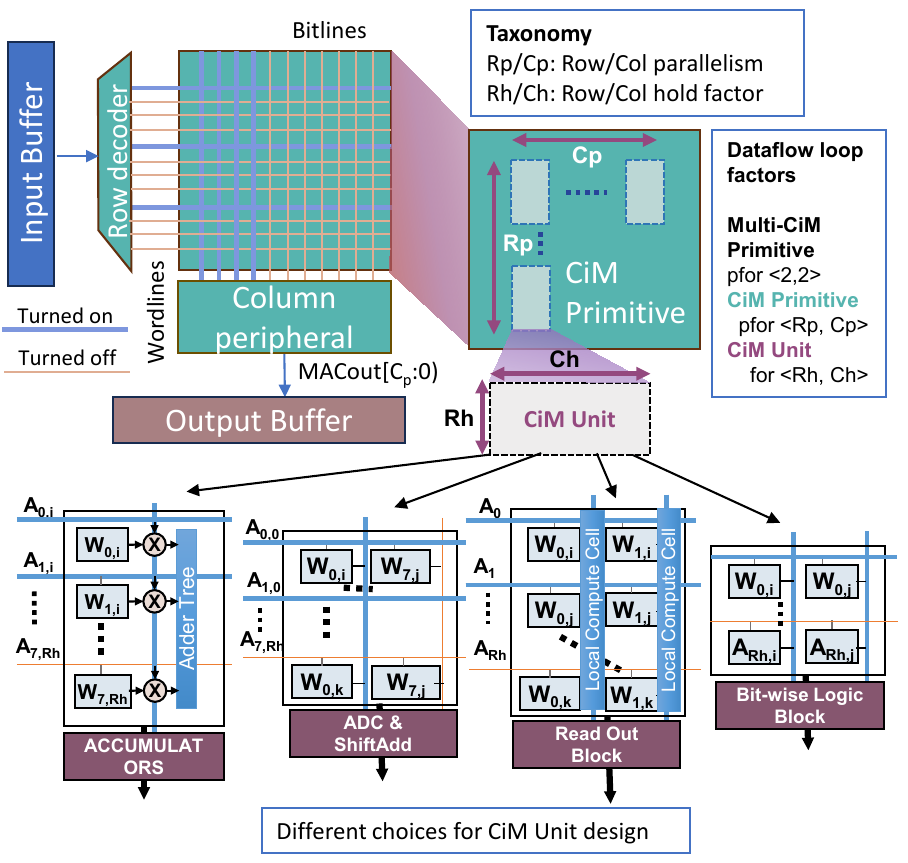} 
  \caption{Diagram showing a typical compute-in-memory primitive consisting of input buffer for feeding multiple wordlines and output buffer for accumulating the MAC results computed across multiple bitlines. Our dataflow-centric approach logically exposes the CiM primitive in terms of parallel $Rp \times Cp$ CiM Units, each capable of performing $Rh \times Ch$ MAC operations sequentially. The physical design of CiM Unit is dependent on technology choices and circuit design. The diagram shows 4 typical choices, including analog and digital primitives.}
  \label{fig:cim-primitive}
\end{figure}

  In this section, we describe our dataflow-centric approach to represent CiM primitives that is used in all subsequent evaluations. The term CiM primitive is used to refer to a CiM array, or in other words, an SRAM array modified to perform in-situ MAC operations. A CiM primitive further consists of multiple CiM units as explained in the first sub-section that provides details on how CiM is integrated in a cache memory. Next, we provide details on the energy and latency estimation of a CiM primitive, followed by the mapping algorithm used to schedule GEMMs on a CiM integrated architecture.

  \subsection{CiM representation}
  Cache memory is generally divided into multiple banks, which can be accessed in parallel. Each bank further contains multiple sub-arrays, accessed one at a time, to reduce wordline and bitline delays. Cache integrated with CiM would replace the standard banks with multiple CiM primitives, especially designed for high energy efficiency. Each CiM primitive is capable of performing multiply-and-accumulate (MAC) operations in parallel. Other than the technology and circuit-level differences, the main architectural difference between standard arithmetic units and CiM primitives is that of stationarity. CiM primitives help in reducing the data movements by keeping data stationary in the memory array and reusing it to perform MAC operations. Owing to this difference in memory accesses, we follow a dataflow-centric approach to represent CiM primitives.

  A typical CiM primitive is shown in Fig. \ref{fig:cim-primitive} where multiple wordlines (rows) and bitlines (columns) can be turned on simultaneously. Generally, rows and columns in a CiM primitive are activated in a time-multiplexing manner due to various limitations. For example, analog CiM primitive based on SRAM-6T requires staggered row activation\cite{ali2020imac} or needs local computing cells to avoid read disturb issues{\cite{analog1cim}}. Further, to reduce the signal-to-noise ratio or compensate for the overhead of ADCs in analog CiM, row decoder activates fewer wordlines at a time or share the same ADC across multiple bitlines. On the other hand, digital CiM primitives, where inputs and weights are stored in the same column, only activates two rows to perform bit-wise operations. We refer to these inherent limitations in row and column parallelism within a CiM primitive in terms of $R_h$ (row \textit{hold}) and $C_h$ (column \textit{hold}) factors, respectively. In contrast, $R_p$ and $C_p$ capture the number of \textit{parallel} MAC operations happening within each primitive. Thus, each CiM primitive is divided into $R_{p} \times C_{p}$ CiM units, where all CiM units can operate in parallel. Here, each CiM unit can \textit{sequentially} perform $R_{h} \times C_{h}$ MAC operations. Our approach can be easily extended to different types and number of CiM primitives and successfully captures the compute parallelism of CiM primitives.

  \subsubsection{CiM cost estimation}
  To incorporate the technological and circuit-level implications of CiM primitives, we follow an approximate approach by taking the costs of energy, latency and area directly from the prototypes. Since the prototypes differ in terms of supply voltage and technology, the energy numbers are scaled to match 45nm technology with 1V supply according to the established work on technology scaling \cite{stillmaker2017scaling}. The difference in frequency of operation of the CiM primitives is captured through compute cycles in terms of latency, by assuming 1GHz frequency of operation. The following equations describe the calculation of energy, latency and area overhead:
\begin{equation}
      Compute~energy~(pJ/mac) = \frac{2}{TOPS/W}*T_{ratio}
  \end{equation}

  \begin{equation}
      T_{ratio} = f_{45nm}/f_{ref}
   \end{equation}

  \begin{equation}
      f_{45nm} = a_{e2}^{45nm}+a_{e1}^{45nm} + a_{e0}^{45nm} 
  \end{equation}
   
   \begin{equation}
      f_{ref} = a_{e2}*V^2+a_{e1}*V + a_{e0} 
  \end{equation}

Here, $a_i$ are unique factors that depend on the technology node \cite{stillmaker2017scaling} \footnote{$a_{e2}^{45nm}$, $a_{e1}^{45nm}$ and $a_{e0}^{45nm}$ are equal to 1.103, -0.362 and 0.2767 respectively.} and $V$ is the supply voltage used in the reference design.
 \begin{equation}
      Compute~latency~(ns) =  \frac{1~GHz}{CiM~frequency} * Cycles_{mac}
  \end{equation}
    \begin{equation}
      Area~overhead(\times) = \frac{Reference~CiM~area}{SRAM~area_{iso-capacity}}
  \end{equation}

Note that, the compute energy here considers the energy of all the components comprising a CiM primitive, including the energy from the cell array, analog-digital converters (ADCs), digital-analog converters (DACs), register for bit-streaming the inputs to the array (input driver), row/col decoder, shift-and-add block, digital arithmetic logic, multiplexers, buffer for temporarily storing the MAC output and any specialized circuit block required for CiM functionality. Thus, our approach strikes the right balance between fully accurate, rigid hardware simulation and low accuracy, flexible software modelling. 

\begin{algorithm}[t]
\caption{Dimension Optimization for N}\label{alg:capopt}
\begin{algorithmic}
\Require Capacity, $M_{used}$, $N_{used}$, $K_{used}$\\
\State $A_{size} \gets K_{used} * M_{used}$ \Comment{Input partition size}
\State $Z_{size} \gets N_{used} * M_{used}$ \Comment{Output partition size}
\State $factor \gets 1$

\While{$A_{size} + Z_{size} \leq Capacity$} \Comment{Check capacity}
     \State $N_{factor} \gets Min factor(N_{rem} // factor)$
     \If{$N_{factor}$ is not None}
         \State $Z_{size} \gets N_{factor} * factor * Z_{size}$
        \State $factor \gets factor * N_{factor}$
    \Else
        \State break \Comment{N dimension is fully mapped}
    \EndIf
 \EndWhile\\

 \Return factor \Comment{Maximum $N_{factor}$ for this memory level}
\end{algorithmic}
\end{algorithm}




\begin{figure}[!t]
    \centering
    \includegraphics[width=\columnwidth]{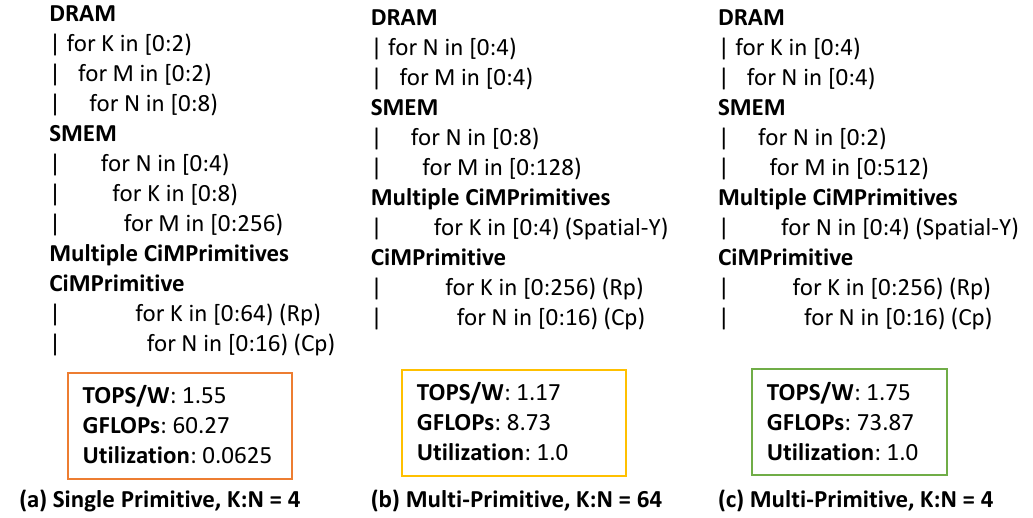}
    \caption{Different dataflow choices for an architecture consisting of 4 fully-parallel CiM primitive type\cite{digitaltsmc}, each having $R_p$ and $C_p$ as 256 and 16, respectively. (a) displays high input reuse while low utilization. (b) shows high utilization but lower throughput due to increased data accesses. (c) shows optimal dataflow with high utilization and high input reuse.}
    \label{fig:dataflow-exdiff}
\end{figure}



\subsection{CiM Mapping}
  Given a CiM primitive and GEMM shape, it is important to efficiently allocate and schedule the MAC operations by choosing an optimal dataflow. One method to find the optimal dataflow is using heuristic search which iterates over different options of mapping before meeting a victory condition\cite{parashar2019timeloop}. However, such search algorithm is agnostic of the inherent reuse opportunities present in a CiM primitive. To that end, we devise a priority based algorithm for deciding the mapping constraints given the CiM architecture and GEMM shape. Our mapping algorithm prioritizes high utilization of CiM units while maintaining weight stationary dataflow, followed by optimizing reuse and reducing data movement from main memory. 

  \textbf{Deciding loop factors}: The first priority is to keep the weight stationary to reduce number of writes to the memory associated with CiM units. This results in mapping the reduction dimension ($K$) to the rows and output columns ($N$) to the columns of all the CiM primitives. Then, the remaining weight dimensions are mapped based on the hold factors, to maintain high utilization of CiM primitives. This ensures that the output partial sums are kept stationary during row hold. The second priority is to maximise the input\//weight reuse by mapping the maximum possible input matrix ($M \times K$) to the adjacent memory level, say SMEM. To achieve this, highest factor of $M$ is calculated which can satisfy the memory capacity constraints. Then, the $K$ and $N$ dimension are optimized for this level, by incrementally increasing the factors. The dimension optimization algorithm for $N$ is described in Algo. \ref{alg:capopt}. The loop factors for the next memory level can be found similarly. The algorithm assumes that the last memory level, DRAM, is large enough to fit all the matrices. Fig. \ref{fig:dataflow-exdiff}(a) illustrates a sample mapping for GEMM($512,1024,1024$) and an architecture integrating CiM primitives at register file (RF).

  In case of multiple CiM primitives, priority is given to higher parallelism (using multiple CiM primitives simultaneously) than fully utilizing each CiM unit. Hence, weights are first mapped to multiple primitives and then to sequential rows and columns of a CiM unit. This could result in a weight-interleaved mapping if the weight size is small enough to fit without using the sequential rows/columns of CiM unit. Further, to decide whether to expand in $K$ or $N$ across CiM primitives, we follow a simple rule that the ratio of larger dimension to smaller dimension should be less than a threshold ($= 4$ for our experiments). We decide the value of this threshold based on our experimental observations, which demonstrated increased memory accesses for higher threshold. Fig. \ref{fig:dataflow-exdiff}(b) shows how the performance and energy efficiency can deteriorate for a skewed (high threshold) mapping compared to a balanced mapping in (Fig. \ref{fig:dataflow-exdiff}(c)) across multiple CiM primitives (or arrays). Multi-CiM primitive mapping can be expanded in future to also include weight duplication, that is, mapping $M$ across primitives. 
  
  \textbf{Deciding loop order}: To leverage the inherent temporal reductions of CiM arrays, the loop order for compute remains as $M < K < N$, where M is in the innermost loop for highest input reuse. By changing $K$ faster than $N$, we prioritize reducing the output partial sums in the CiM primitive before moving to different partial sum. For other memory levels, priority is given to reducing the number of data accesses using a greedy algorithm. The greedy approach tries to minimize the number of data accesses locally for a memory level in the hope of reaching a globally optimal solution. Since the dimension in the outermost loop affects the number of data accesses for other dimensions in a memory level, the algorithm keeps the final loop order such that the dimension with largest loop factor is kept in the outermost \textit{for loop}. For instance, in Fig. \ref{fig:dataflow-effect} (a) where $M1=3$ is in outermost dimension, all access factors are multiplied by $3$ and similarly in Fig. \ref{fig:dataflow-effect} (b) where $K1=2$ is outermost loop, all access factors have $2$ as the common factor. So, by keeping the lowest dimension in the outermost loop, we minimize the access factors locally (greedily) at each memory level.

\begin{figure}[!t]
  \centering
  \includegraphics[width=\columnwidth]{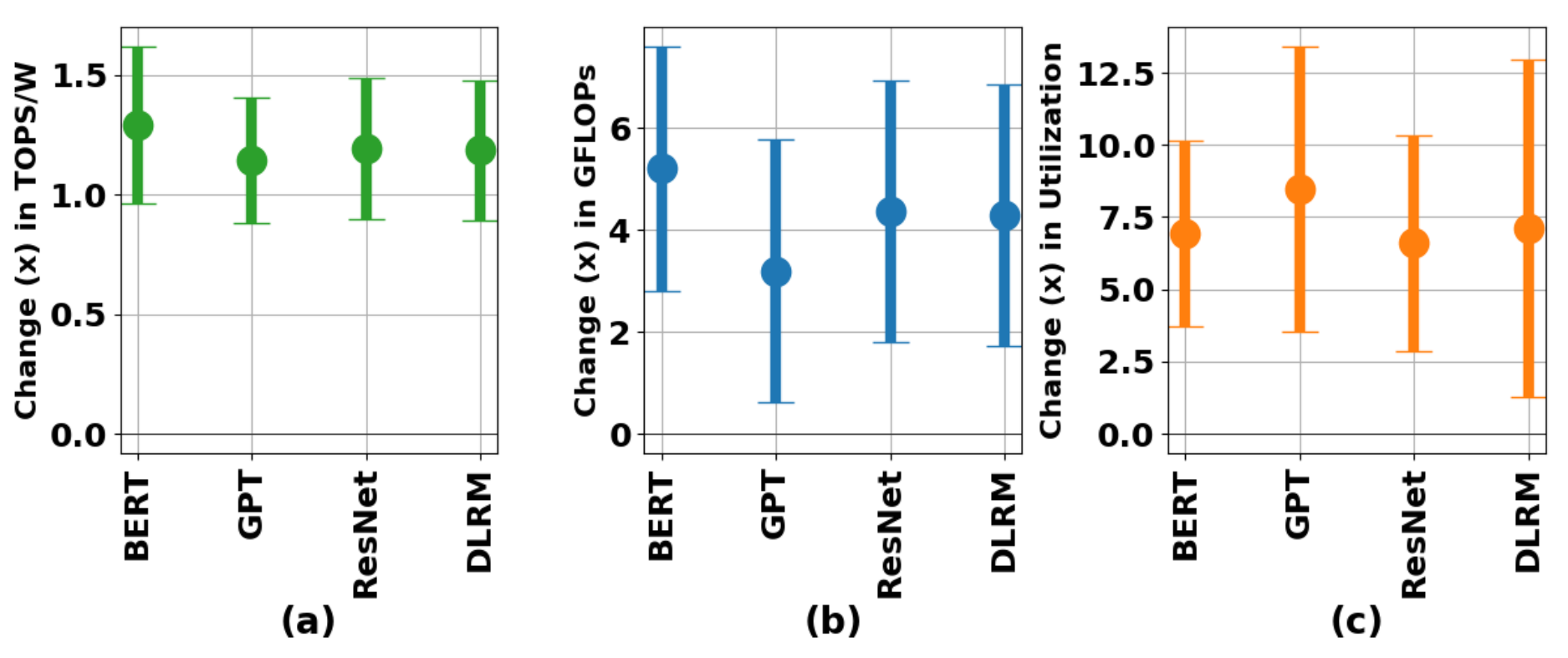} 
  \caption{Error bars showing average speedup and standard deviation in speedup for (a) TOPS/W, (b) GFLOPS and (c) Utilization over a variety of GEMM shapes for a typical digital CiM primitive. Change $ > 1$ shows that our mapping optimization strategy performs better than heuristic search which stops after encountering 100,000 consecutive invalid mappings.}
  \label{fig:algo-results}
\end{figure}

\textbf{Comparison with Heuristic Mapping}:
Fig. \ref{fig:algo-results} shows the difference in energy efficiency ($TOPS/W$), performance ($GFLOPS$) and hardware utilization across different GEMM shapes, when compared with heuristic mapping search. The circles in the error bars represent the average change (or speedup) while the length is proportional to standard deviation in speedup. The results prove that our mapping algorithm consistently performs better in terms of utilization of resources and energy efficiency across all the GEMM shapes, including those present in ML workloads such as BERT-Large, GPT-J, ResNet50 and DLRM. As described earlier, our algorithm prioritises high utilization of CiM units by mapping largest possible weight matrix. This is in accordance with the results which show more than $6.6 \times$ improvements in utilization, on an average, across different types of GEMMs. Given the huge mapspace, our algorithm improves energy-efficiency by $1.2 \times$ while enhancing the throughput by more than $3.2 \times$, on average.

Hence, our priority based mapping algorithm provides optimal dataflow mapping constraints given a CiM based primitive and GEMM specification. In addition, Table \ref{tab:algo-time} shows that the runtime cost of generating the mapping constraints using our algorithm is not significant compared to heuristic search and the benefits increase with the number of runs. Moreover, our algorithm always provides a valid mapping, unlike the heuristic search which requires iterative tuning of the search configurations to find the final mapping. Therefore, our consistent mapping strategy helps in developing a better understanding when comparing benefits of CiM integration across different GEMM shapes and architecture specifications. 

\begin{table}[t]
    \centering
    \caption{Comparison of User Runtime (in seconds)}
    \begin{tabular}{|c|c|c|}
    \hline
    Number of runs & Our Algorithm & Heuristic Search \\
    \hline
    5 & 42.41 & 56.17 \\
    10 & 85.90 & 112.53 \\
    50 & 408.12 & 562.46 \\
    \hline
    \end{tabular}
    \label{tab:algo-time}
\end{table}

\section{Experimental Setup}
\label{sec:methodology}

\subsection{Baseline Architecture}
  The baseline architecture consists of a single SM connected to main memory, akin to a GPU processing core, as shown in Fig. \ref{fig:intro}. Here, each sub-core in an SM consists of $16 \times 16$ processing elements (PEs) to perform MAC operations, representing tensor-core-like operations. Note that a GPU consists of hundreds of such SMs, resulting in overall peak performance of the order of PFLOPS, while our baseline architecture consists of only 1 SM. Further, we consider INT-8 precision, 45nm technology and 1GHz frequency of operation in all evaluations. INT-8 is chosen for its acceptable accuracy in ML inference tasks\cite{nagel2019data,dettmers2022llm}. The cache capacity of register file (RF) and shared memory (SMEM) are assumed to be $4 \times 4~KB$ and $256~KB$ respectively. The bandwidth for SMEM and DRAM are taken as $42B$/cycle and $32B$/cycle. The energy cost of different components is displayed in Table \ref{tab:exp-energy}.

\begin{table}[t!]
  \centering
  \caption{Energy cost with INT-8 data precision (45nm)\cite{iccad_2019_accelergy}}
        
  \label{tab:exp-energy}
  \small
  \begin{tabular}{|l|l|}
    \hline
    Component & Energy (in pJ) \\
    \hline
    DRAM access & 512.00 \\
    \hline
    SMEM access & 124.69 \\
    \hline
    RF access & 11.47 \\
    \hline
    PE buffer access & 0.02\\
    \hline
    MAC compute & 0.26\\
    \hline
  \end{tabular}

\end{table}

\subsection{CiM Primitive Choices}

\begin{figure*}[ht]
    \centering
    \includegraphics[width=\textwidth]{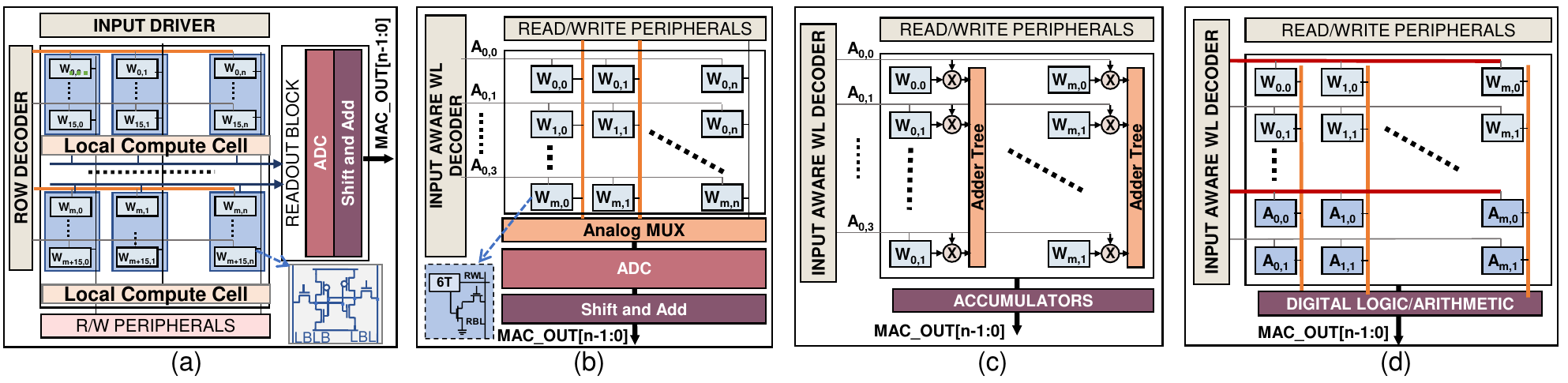}
    \caption{Examples of design choices for integrating CiM primitives in memory with (a) and (b) representing Analog-6T and Analog-8T, and (c) and (d) representing Digital-6T and Digital-8T, respectively.}
    \label{fig:CiM-primitives}
\end{figure*}

\begin{table*}[ht]
    
  \caption{Single CiM primitive Specification}
        
  \label{table:CiM-list}
  \small
  \begin{tabular}{|l|l|l|l|l|l|l|l|l|l|l|}
    \hline
    \multirow{2}{*}{\textbf{Sr.No.}} & \multicolumn{10}{|c|}{\textbf{Specification}} \\
    \cline{2-11}& Compute Type& Cell Type& Rp& Cp &  Rh& Ch& Capacity (KB)&Latency (ns)&8b-8b MAC energy (pJ)&Area (x)\\
    \hline
    1& Analog & SRAM-6T & 64 & 4  & 1 & 16 & 4 & 9 & 0.15 & 1.34\\
    \hline
    2& Analog & SRAM-8T & 64 & 4  & 1  & 16 & 4 & 144 & 0.09 & 2.1\\
    \hline
    3& Digital & SRAM-6T & 256 & 16  & 1  & 1 & 4 & 18 & 0.34 & 1.4\\
    \hline
    4& Digital & SRAM-8T & 1 & 128  & 10  & 1 & 4 & 233 & 0.84 & 1.1\\ \hline
  \end{tabular}

\end{table*}

The chosen primitives cover a range of distinct and varying parameters, as listed in Table \ref{table:CiM-list} and explained below. 

The \textbf{SRAM6T based analog} \cite{analog1cim} CiM primitive, illustrated in Fig. \ref{fig:CiM-primitives}(a), employs local computing cells (LCCs) shared within a block of rows storing weights. LCC facilitates MAC operations between inputs passed through the input driver and the weights stored in 6T cells. Then, the analog output from LCC goes through a readout block including ADC and shift-and-add circuitry to obtain the final bits of MAC output. Here, different input bits can be fed in parallel to multiple columns, resulting in lower compute latency. However, the number of LCC blocks and ADCs limit the compute parallelism of the CiM primitive.

An example of \textbf{SRAM8T based analog\cite{ali2023cicc} primitive} is depicted in Fig. \ref{fig:CiM-primitives}(b). It can enable different number of rows (wordlines) in the array depending on the input sparsity. Such re-configurable ADC design offers high energy-efficiency, but suffers from relatively higher peripheral cost (area overhead) due to bigger ADC design. Additionally, feeding inputs in a bit-serial fashion to the same row results in a higher compute latency for such design. 

The \textbf{digital\cite{digitaltsmc} primitive based on SRAM6T cells} is shown in Fig. \ref{fig:CiM-primitives}(c), that employs a fully digital design, feeding inputs into each row and executing a MAC operation at every cross-point, later combined using an adder tree. Here, the adder tree reduction incurs an area overhead and results in a higher compute latency than its analog counterpart.

One of the first \textbf{digital primitives based on SRAM8T cells}, shown in Fig. \ref{fig:CiM-primitives}(d), displays a design where both inputs and weights are mapped to the same column. This configuration allows only two rows to be turned on at a time for a 1b-1b bitwise operation. Due to the heavily bit-serial nature of MAC computation in this design, is takes the highest compute latency but offers a minimal area overhead.


\subsection{Workload Choices}

To cover a range of shapes with different memory and compute requirements, we consider a blend of synthetic and real data. Synthetic GEMMs dataset consists of $1000$ datapoints with $M$, $N$ and $K$ varying from $16$ to $8192$. Real dataset consists of GEMM shapes from different ML models, assuming single batch size during inference. We characterized ResNet50 \cite{he2015deep} with ImageNet \cite{deng2009imagenet}, BERT-Large \cite{devlin2018bert} with sequence length equal to 512, DLRM \cite{naumov2019deep} and GPT-J decoding phase\cite{gpt-j} to create the real dataset.

\subsection{Evaluation Metrics}
The total energy of the system is the weighted sum of the individual arithmetic cost, memory access costs and reduction costs multiplied by the number of arithmetic operations, memory accesses and reductions. The energy cost per MAC operation is dependent on the CiM primitive while the temporal reduction cost per addition is considered to be a constant at $0.05pJ$. Subsequently, the number of memory accesses and reductions is highly dependent on the dataflow. Given the dataflow, $TOPS/W$ (Tera Operations Per Second per Watt) can be calculated as the ratio of number of operations by total energy cost. For throughput calculation, we assume a fully pipelined system with compute and memory accesses overlapping with each other. The total number of cycles is calculated by taking the maximum of compute cycles and memory cycles, which are again dependent on the dataflow. Here, a compute cycle assumes a CiM primitive having pipelined operations of input buffer read, MAC computation and output buffer write, and that the weight loading latency is hidden in the compute cycles. The total number of cycles help in determining the throughput of the architecture, where $GFLOPS$ (Giga Floating Point Operations per Second) is the ratio of total number of operations by the total cycles. Further, utilization refers to the compute hardware utilization in our results, such that it is equal to the number of utilized MAC units by the total number of MAC units. Note, that each CiM unit consists of $R_{h} \times C_{h}$ MAC units by definition.

\section{Results and Discussion}
\label{sec:results}
 The purpose of this section is to address the fundamental questions posed at the beginning of the paper regarding the optimal conditions for utilizing SRAM-based CiM primitives -- specifically what, when and where to integrate them. Note, that all the results in this section adhere to the iso-area constraints, ensuring that the on-chip cache area remains the same after CiM integration. Thus, the number of CiM primitives (consequently the memory capacity) that can be accommodated in cache depends on the area overhead of the CiM primitive choice. Further, the graphs correspond to CiM integrated in register file (RF) unless stated otherwise.

\subsection{Effect of CiM Primitive choice}
\begin{figure}
    \centering
    \includegraphics[width=\columnwidth]{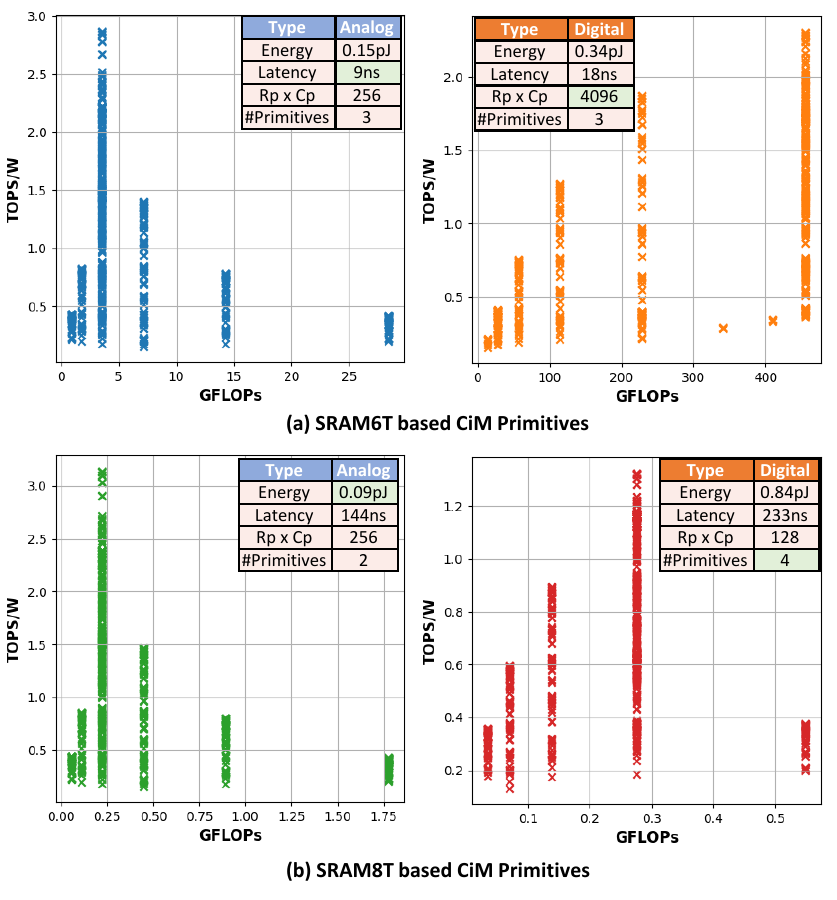}
    \caption{Energy-efficiency vs Throughput with different primitives for an architecture having CiM integrated at register level under iso-area constraints. (a) shows SRAM6T based analog\cite{analog1cim} and adder tree digital \cite{digitaltsmc} primitives, and (b) shows SRAM8T based re-configurable ADC analog\cite{ali2023cicc} and bitwise logic digital \cite{computesram} primitive performance. The area overhead of each primitive decides the number of CiM primitives that can fit within the memory area. Each datapoint corresponds to a $GEMM(M,N,K)$ where $M,N,K$ varies from $16~to~8192$.}
    \label{fig:results-what}
\end{figure}

 Fig. \ref{fig:results-what} compares the energy-efficiency of the chosen CiM primitives for different $GEMM(M,N,K)$ shapes from the synthetic dataset (where $M,N,K$ varies from $16~to~8192$). The graphs in Fig. \ref{fig:results-what} indicate that the CiM primitive with lowest energy cost ($0.09pJ$) also shows highest energy efficiency at system level, achieving more than $3~TOPS/W$. However, the advantage in energy efficiency (by $1-2~TOPS/W$) in an architecture including DRAM is significantly lower than standalone CiM accelerators which showcase two to three orders of magnitude in TOPS/W\cite{digitaltsmc}. This highlights the critical bottleneck -- while CiM can dramatically increase energy efficiency, the benefits are considerably constrained by the slower, more energy-intensive main memory accesses.

Furthermore, the analysis reveals that the inherent sequential processing nature of CiM primitives, due to mechanisms like row and column multiplexing, limits the overall system throughput. For example, while SRAM6T based analog design has lower latency than its digital counterpart, analog design cannot fully compensate for the reduced parallelism compared to digital. Our results indicate that SRAM6T based digital design, which includes adder trees for MAC computation, achieves the highest throughput in general. On the other hand, SRAM8T based digital CiM with bitwise logic peripherals underperforms in throughput compared to other architectures, despite its lower area. These findings emphasize the trade-off between latency, parallelism and area-overhead in CiM architectures. 

\textbf{Key Takeaways:} The digital CiM primitive with an adder tree design provides an optimal trade-off between throughput and energy efficiency, proving effective for balanced performance requirements. Analog CiM primitives obtain a better signal-to-noise ratio and saves area overhead though row and column multiplexing. However, such techniques heavily hinder the overall system performance. Hence, there is a need to design analog CiM primitives with lower latency to leverage the high energy efficiency of analog design while maintaining performance. One possible option is to have ADC-less analog CiM designs which can eliminate the high latency and area overhead of bulky ADCs\cite{saxena2022towards}.

\subsection{Effect of Workload choice}

\begin{figure*}[t]
    \centering
    \includegraphics[width=\textwidth]{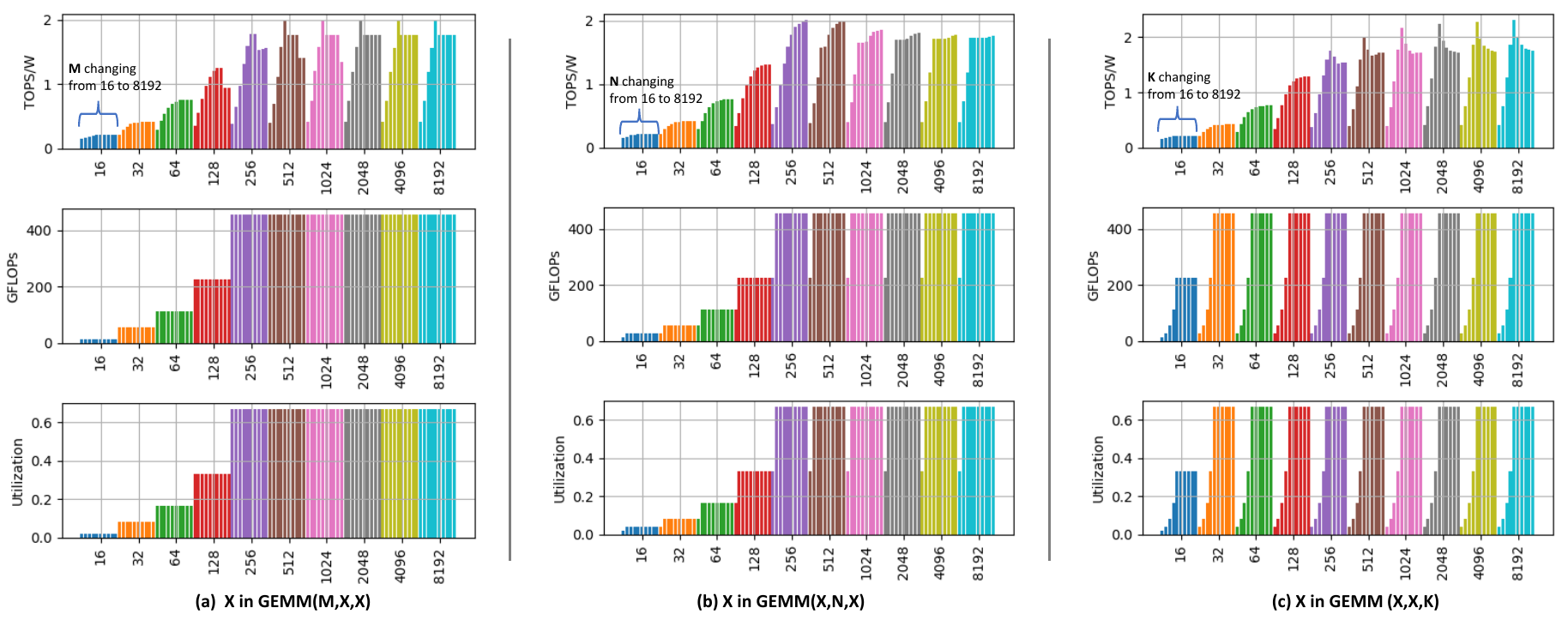}
    \caption{Comparison of change in energy-efficiency, throughput and utilization for a typical digital CiM primitive (Digital6T) integrated in register file. Each sub-graph shows how the evaluation metrics change with a particular dimension of GEMM, when other two are kept fixed at X. Particularly, (a) shows impact of weight matrix ($N \times K$) and change in $M$ dimension for each matrix size, (b) shows impact of input matrix ($M \times K$) and change in $N$ dimension, and (c) shows impact of output matrix ($M \times N$) and change in $K$ dimension for a fixed output matrix size.}
    \label{fig:results-when}
\end{figure*}

As observed in Fig. \ref{fig:results-what}, the energy-efficiency and performance of CiM primitives is significantly influenced by the dimensions of the matrices involved in GEMM operations. This section presents a detailed analysis of how variations in the dimensions of weight, input, and output matrices influence key performance metrics, such as TOPS/W, GFLOPS, and hardware utilization. 

\textbf{Impact of weight matrix:} Fig. \ref{fig:results-when}(a) shows how the key performance metrics change with the regular shaped weight matrix ($N = K$). For each weight matrix size, the input shape is varied by changing $M$ and hence varying reuse opportunities. As shown in the TOPS/W  sub-figure, the energy efficiency of CiM primitive based architecture progressively increases from $N = K = 16$ until dimension $K$ reaches the on-chip memory capacity, in this instance capped at a value of $512$. Beyond this threshold, energy efficiency stabilizes at approximately $1.75~TOPS/W$, showing no further gains despite increases in the weight matrix size. This plateau is primarily due to the proportional increase in access energy cost with the weight size. Once the weights are fully mapped, further increase in the weight size results in re-writing of weights to the CiM primitives from the main memory, limiting the maximum on-chip reuse and hence the constant energy efficiency.

For a fixed size of weight matrix, as $M$ increases, the reuse opportunities increase. Hence, energy efficiency can range from $\approx~0.3~TOPS/W$ to $1.9~TOPS/W$ with the change in input shape. However, this trend is followed only to a sweet point of $M$. For instance, with a matrix size of $512 \times 512$, TOPS/W dramatically falls from $1.97$ to $1.75$ as $M$ grows from $256$ to $512$. The primary reason behind the adverse effect on performance is the increased accesses to the main memory once $M$ exceeds the shared memory capacity constraints. Another important observation from the graph is that the TOPS/W is limited by the $M$ dimension across different weight matrix sizes. For example, for $M = 32$, energy efficiency does not go beyond $0.73~TOPS/W$ irrespective of the weight matrix size.

On the other hand, throughput follows a simpler trend and continues to rise with weight matrix size until it reaches the full utilization point where all CiM units are working in parallel. When 2 out of 3 CiM primitives are used, corresponding to $K = 256$ and $N = 32$ in this case, it achieves the maximum throughput of $455$GFLOPS. Utilization depends on the size of the GEMM shape and maximum compute capacity of CiM primitives. Note, that the throughput does not change with $M$ because there is no bandwidth throttling from memory for $M > 16$. Generally, the compute cycles from CiM primitives exceed the memory access cycles due to their operation on lower frequency.

\textbf{Impact of input matrix:} Fig. \ref{fig:results-when}(b) shows how the key performance metrics change with varying input matrix ($M = K$) and varying $N$ for a fixed input matrix size. The relationship between the input matrix size ($M \times K$) and the $TOPS/W$ metric initially shows an increase as the matrix dimensions expand. However, this growth in $TOPS/W$ begins to taper off once the matrix size reaches and surpasses the $M = K = 512$ threshold. Fig. \ref{fig:results-when}(a) highlighted that increase in $M$ does not affect the energy efficiency beyond a certain threshold. Hence, this tapering in energy efficiency is attributed to increase in $K$, as also evident in Fig. \ref{fig:results-when}(c).

Interestingly, the increase in the size of the $N$ dimension positively affects energy efficiency, which continues to rise irrespective of the size of the input matrix. The rate of increase in energy efficiency reduces with extremely large matrices ($> 4096$). Since $N$ is the common dimension in weights and outputs, this suggests that the weight stationary nature of CiM primitives helps in efficiently leveraging the output reuse and minimizing data accesses with increase in output matrix width. 

Throughput also responds significantly to changes in the size of the N dimension. When $N$ is set to 16, only a single CiM primitive is engaged, which caps the throughput even for larger matrix sizes. As $N$ increases, the architecture leverages more CiM primitives in parallel, thereby enhancing the throughput substantially. 

\textbf{Impact of output matrix:} Fig. \ref{fig:results-when}(c) shows the change in key performance metrics with output matrix ($M = N$) and varying $K$ for a fixed input matrix size. Note, that $K$ is the reduction dimension in the GEMM operation. As the dimensions of the output matrix ($M \times N$) increase, a corresponding rise in $TOPS/W$ is observed, albeit with a diminishing rate of increase as the matrix size enlarges. This deceleration in energy efficiency highlights the challenges in scaling up the output dimensions without encountering diminishing returns in terms of $TOPS/W$.

For the dimension $K$, $TOPS/W$ initially rises as the dimension increases, reaching an optimal point when $K$ is equal to $256$. This \textit{sweet spot} corresponds to the maximum number of rows ($256$) that the digital CiM primitives can process simultaneously, optimizing the balance between computation and memory access. Beyond this point, further increases in $K$ result in a decline in $TOPS/W$. The decrease is primarily due to the increase in the number of output partial sum accesses from main memory, which adversely impacts the overall system performance by increasing the data traffic and energy consumption.

In contrast, throughput consistently increases with the expansion of both N and K dimensions, up to a point where all CiM units are fully utilized. Once this saturation point is reached, throughput plateaus. This pattern of utilization mimics the throughput trend, indicating a direct correlation between the utilization of computational resources and the achieved throughput.

\begin{figure*}[!ht]
    \centering
    \includegraphics[width=0.9\textwidth]{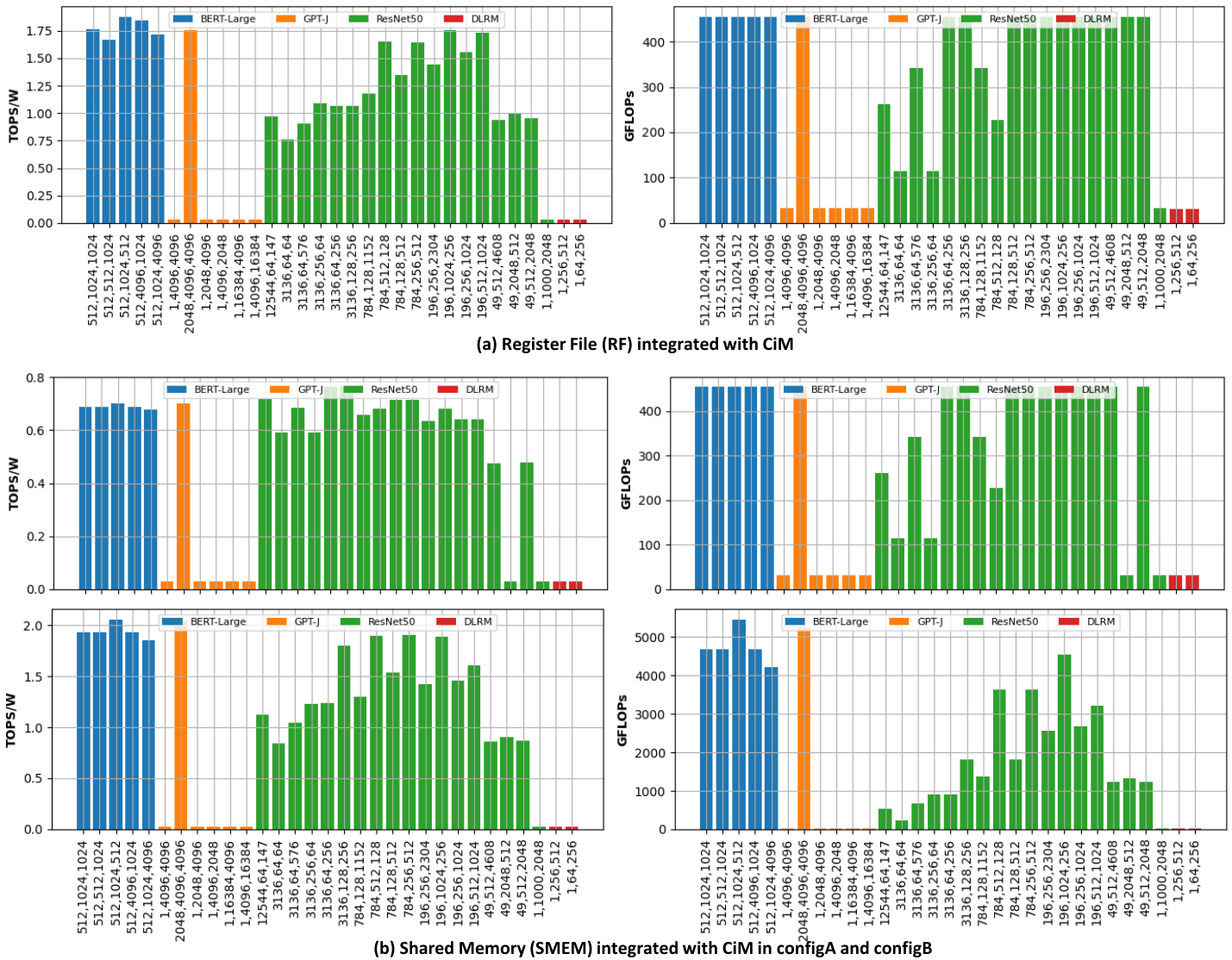}
    \caption{Energy-efficiency and performance for GEMMs based on different ML workloads when integrating Digital-6T CiM primitive at (a) register file (RF) and (b) shared memory (SMEM) level under iso-area constraints. configA considers an equal number of CiM primitives as the RF integration for compute. configB includes all CiM primitives that can be accommodated within the existing memory capacity of SMEM (16x compared to configA).}
    \label{fig:results-where}
\end{figure*}

\textbf{Key takeaways: } The effect of changing N, K and M dimensions is different on CiM energy efficiency benefits. While the energy efficiency always increases with increase in N, it diminishes when K is increased beyond the value that can be reduced in the CiM primitives itself. For the M dimension, energy efficiency rises till a certain value when all the inputs can be mapped to the shared memory present on chip.
One common trend among all dimensions is that, irregular shaped GEMMs (where one dimension is much smaller than other dimensions) perform poorly with CiM primitives in terms of both energy efficiency and throughput. On the other hand, large regular shaped GEMMs achieve almost equal optimal energy efficiency. Here, such shapes are at least $4$ times bigger than what the CiM primitives can store in the memory ($GEMM(X,X,X)$ where $X > 1024$). Highest energy benefits are achieved when the weight matrix size is fully mapped to the CiM storage capacity and input is fully stored in the shared memory.

Utilization for CiM primitives is directly influenced by the sizes of the N and K dimensions, due to their weight stationary nature. Hence, maximum throughput is achieved when the weight matrix size is sufficiently big to utilize all CiM primitives. Throughput does not diminish with change in $M$ dimension, primarily because CiM primitives have high compute cycles, mitigating potential bandwidth throttling issues from memory. However, if reuse is extremely low (next-subsection), it could lead to lower throughput due to memory bandwidth throttling.

\subsection{Effect of Memory Level}

Having established how various GEMM shapes impact CiM performance, our subsequent analysis utilizes real datasets representative of typical ML inference workloads to understand where to integrate CiM in the on-chip memory hierarchy. We investigate CiM integration at two memory levels within a tensor-core like architecture: the register file (RF) and the shared memory (SMEM). The integration of CiM in SMEM is analysed using two configurations for better understanding of the memory hierarchy benefits. Configuration A maintains an equivalent number of CiM primitives as the RF integration, preserving parity in computational resources. Configuration B, in contrast, expands the utilization to include all CiM primitives that can be accommodated within the existing memory capacity of SMEM. This approach allows us to assess the trade-offs and benefits of integrating CiM within the constraints of on-chip memory area.

\textbf{Register File:} Energy efficiency and performance benefits from integrating CiM at the RF level vary across different ML workloads as shown in Fig. \ref{fig:results-where}(a). BERT-Large layers achieve exceptional energy efficiency ($> 1.67~TOPS/W$) and the highest throughput ($455 GFLOPS$) due to their large, regular-sized GEMM shapes. GEMM shape from the GPT-J decoding phase, which is part of the feed-forward layer, exhibits a similar trend due to its large, regular size. The benefits for ResNet-50 layers, however, are more variable, heavily influenced by GEMM shape. The effect of GEMM shape on the evaluation metrics was explained in the last sub-section. For other layers within the GPT-J decoding phase and DLRM embedding layers, which have $M = 1$, the energy efficiency is as low as  $0.03~TOPS/W$ with correspondingly low throughput, $\approx 31~GFLOPS$. These are extreme cases of irregular-sized GEMM shapes, typically called matrix vector multiplications, and suffer from low data reuse opportunities. Here, both the energy and total cycles are dominated by the main memory accesses, resulting in heavy bandwidth throttling from main memory.

\textbf{Shared Memory:} At the SMEM level, integrating the same number of CiM primitives as at the RF level yields throughput improvements similar to the RF as shown in configA of Fig. \ref{fig:results-where}(b). However, this configuration results in notably lower energy efficiency compared to the RF (upto only $0.70~TOPS/W$). The absence of an intermediate on-chip memory level leads to increased accesses to main memory when integrating CiM at SMEM. In contrast, configB at the SMEM level, which utilizes all the CiM primitives that can fit in SMEM under iso-area constraints, overcomes this limitation. ConfigB significantly enhances the throughput, exceeding RF performance by approximately tenfold. The bigger variation in throughput for configB compared to configA, particularly for ResNet50, is due to the difference in utilization. Additionally, this configuration achieves higher energy efficiency by $\approx 0.25~TOPS/W$ compared to RF. This configuration accommodates larger weight matrix shapes and minimizes duplicate accesses to main memory, improving energy efficiency. Nevertheless, the peak energy efficiency is still limited by the size of $K$ that can be mapped to the CiM primitives. It is also notable that matrix vector multiplication layers exhibit no improvement in energy efficiency, even with an increased number of CiM primitives.

\begin{figure}[t]
    \centering
    \includegraphics[width=0.9\columnwidth]{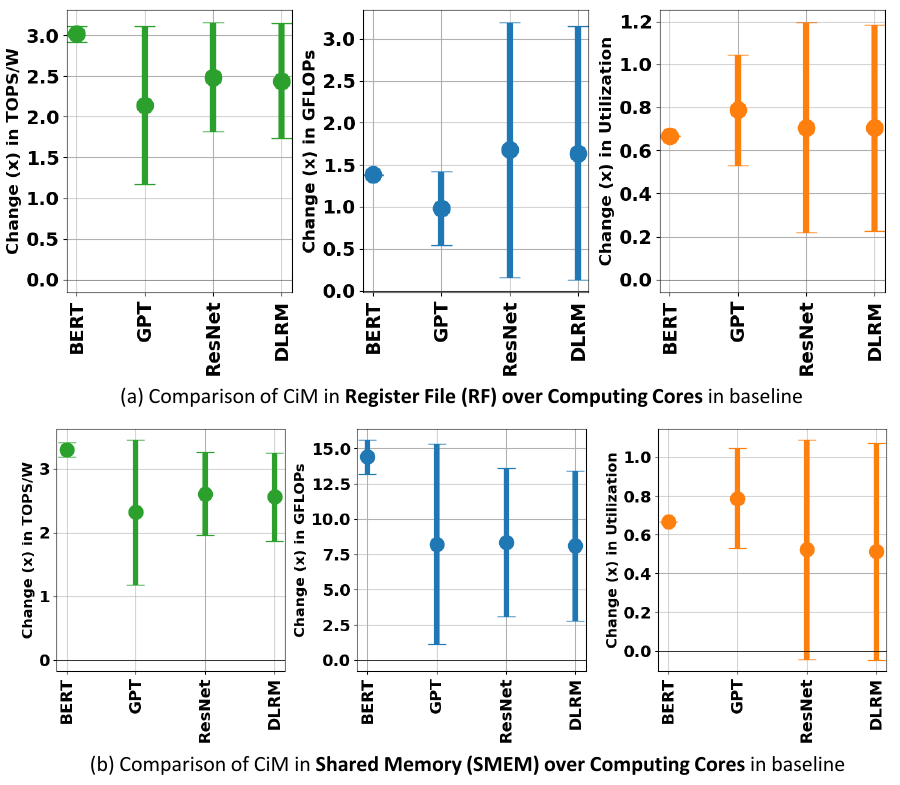}
    \caption{Comparison in TOPS/W, GFLOPS and Utilization when integrating Digital-6T CiM at (a) RF and (b) SMEM relative to tensor-core-like baseline. The round dot shows the average change and the length corresponds to standard deviation in change across different GEMM shapes for the given ML workload. Change $ > 1$ implies that CiM integrated architecture performs better than baseline tensor-core-like architecture.}
    \label{fig:results-avg}
\end{figure}

\textbf{Comparison with baseline:} On average, Fig. \ref{fig:results-avg}(a) compares RF integrated with CiM to a standard or baseline tensor-core architecture. As observed from the graph, BERT layers consistently derive the highest benefits from CiM integration compared to the baseline, with a $3 \times$ increase in $TOPS/W$. While other ML workloads such as ResNet, GPT-J and DLRM also obtain high $TOPS/W$ and $GFLOPS$, some layers exhibit modest or negative changes ($< 1$) relative to the baseline. The underlying reason is the difference in dataflow between tensor-core like architecture and CiM integrated architecture. Since CiM employs a weight-stationary mapping, it heavily relies on the $M$ dimension to leverage weight reuse opportunities. Thus, CiM integration is particularly disadvantaged when $M$ dimension is extremely small. In contrast, tensor-core like architecture does not necessarily maintain weight stationary. This flexibility allows it to better utilize the hardware with even small $M$ cases, capitalizing on other reuse opportunities and achieving a higher throughput than CiM integrated architecture. For bigger GEMM shapes where both architectures can equally leverage the data reuse opportunities, CiM integrated architectures perform better because of higher peak throughput compared to baseline. In terms of energy efficiency, CiM primitives consistently achieves higher $TOPS/W$ than the baseline by saving the data accesses in the lower memory levels.

Fig. \ref{fig:results-avg}(b) plots the difference in evaluation metrics between SMEM and baseline, assuming both memory levels utilize all the CiM primitives that can fit within the memory. The storage capacity of SMEM is $16\times$ bigger than the total RF capacity, which is also reflected in the throughput gains, even when not all CiM primitives are utilized at the SMEM level. The results align with individual GEMMs performances, as illustrated in Fig. \ref{fig:results-where}. 

\textbf{Key Takeaways:} The storage capacity of memory is more important than its hierarchical level when integrating CiM. A higher memory capacity would lead to increased performance due to greater parallelism and slightly improved energy benefits by reducing duplicate memory accesses. However, there are no benefits of CiM integration with ML workloads consisting solely of matrix vector multiplications (GEMMs with $M = 1$). For such cases, integrating CiM at RF performs worse than the baseline tensor-core like architecture unlike integrating CiM at SMEM. On the other hand, SMEM can compensate for the lack of flexibility in mapping by its large compute capability in such scenarios.

\begin{table*}[!t]
    \centering
    \caption{Takeaways}
    \label{tab:key_obs}
    \small
    \begin{tabular}{p{1cm}|p{7cm}|p{7cm}}
    \hline
    Category & Key observation & Reason \\
    \hline
    \multirow{2}{*}{What}
                      &  Maximum throughput gain is achieved by Digital-6T compared to baseline and other CiM primitives for medium to large GEMM shapes.
                      &  The ability to exploit full row and column parallelism in CiM helps in improving the throughput despite higher compute latency given the GEMM shape is large enough to leverage this parallelism.\\
    \cline{2-3}
                      & Analog-8T achieves maximum energy reduction (or highest energy efficiency) compared to baseline and other CiM architectures under is-area constraints.
                      & The overall energy consumption of CiM integrated architectures becomes proportional to their standalone MAC energy when the memory access costs are fully amortized. \\
    \hline
    \multirow{2}{*}{When}
                        & CiM integrated caches do not increase the performance of memory bound layers.
                        & The throughput of memory bound layers is limited by the main memory bandwidth.\\
    \cline{2-3}
                    & GEMMs with high $K$ value have high energy and performance benefits from CiM integration. In contrast, small $K$ GEMMs achieve better throughput with baseline than CiM primitives.
                    & CiM primitives perform in-situ reduction over the K dimension. This reduces the number of memory accesses for partial sum reductions, but puts physical constraints on parallelism across $K$ dimension.\\
    \hline
     \multirow{2}{*}{Where}
                    & Highest performance gains are observed at SMEM level as compared to RF level under iso-area constraints.
                    & Bigger memory capacity allows for more parallelism in CiM integrated architectures and it can help surpass the cache benefits of a less parallel architecture. \\
    \cline{2-3}
                      & The system-level energy-efficiency benefits for SMEM level are slightly higher than RF level.
                      &  For large workloads that can not fully fit data dimensions in the SMEM level, the highly parallel CiM in SMEM can reduce duplicate memory accesses. \\
    \hline

    \end{tabular}

\end{table*}

\subsection{Discussion and Future Work}

Our evaluations are based on analytical modelling of CiM primitives in a tensor-core-like architecture. It assumes all memory accesses are coalesced and does not consider effects such as bank conflicts, limited miss handling buffer capacity, and other architectural optimizations in memory. However, it still captures the approximate performance of different CiM primitives and helps in gaining an understanding of their impact at system level.

  We assumed weight stationary dataflow for all scenarios, as it is the most commonly used and feasible dataflow for CiM primitive. Adding more flexibility in mapping when considering large number of CiM primitives, such as weight duplication, would lead to a bigger design space search and remains an open space to explore in the future. Such an exploration should also take into account the interconnect cost associated with dataflow flexibility. Further, recent works \cite{wu202322nm,yue202328nm} have demonstrated floating point CiM accelerators, expanding the scope of computation in memory. However, all experiments in this work assume INT-8 precision including that in baseline to make it inclusive of a variety of CiM primitives. In addition, we evaluate CiM primitives in terms of performance metrics such as TOPS/W and GFLOPS. The potential accuracy loss associated with analog primitives is not addressed in this work as it falls outside our scope. Note, that CiM integrated architectures would also incur a programming cost overhead that should be considered when finalizing the design. Our work provides early-stage estimations for developing such CiM integrated accelerator architecture designs.

  Based on our evaluations, we offer the following recommendations when integrating CiM in the memory subsystem for machine learning inference:

  \begin{itemize}
      \item CiM offers the best solution for ML inference in terms of energy-efficiency, but lags in performance due to limited parallelism and high latency. Other than CiM primitive choices, dataflow can impact the overall performance and energy-efficiency of CiM architectures, creating the need to have scalable mapping strategies. 
      \item The energy efficiency of CiM-integrated memory subsystem is significantly lower compared to standalone CiM-based accelerators, due to expensive main memory accesses.
      Therefore, to reduce the impact of main memory accesses, designing larger CiM accelerators that can map all matrix dimensions onto on-chip memory is more advantageous.
      \item The impact of cache bandwidth is minimal on CiM integration in most scenarios, due to their low frequency of operation (or equivalently high compute latency). Thus, memory level with highest capacity is the optimal level to integrate CiM. 
      \item The energy efficiency benefits of CiM decrease when the workload dimension $K$ exceeds the reduction capability of the CiM primitive due to increased partial sum accesses. Consequently, the size of CiM primitive-based accelerators should be tailored to accommodate workload-specific reductions in dimension $K$.
      \item Avoid deploying CiM primitives for matrix-vector multiplication tasks where data reuse is minimal.
  \end{itemize}

\section{Conclusion}
\label{sec:conclusion}
In conclusion, our study establishes a foundation for integrating on-chip memory with SRAM-based Compute-in-Memory (CiM) primitives to efficiently perform general matrix multiplications (GEMMs) during machine learning (ML) inference. Leveraging the weight-stationary compute nature of CiM primitives, we introduced a dataflow mapping algorithm tailored towards optimizing the energy-efficiency and performance of CiM integrated architectures. The algorithm prioritizes maximizing weight reuse and minimizing partial output sum accesses for a given CiM primitive and GEMM configuration. While CiM consistently achieves higher TOPS/W than standard tensorcore based design, our evaluations suggest that CiM architectures are particularly well-suited for encoder based transformer models than other ML models. Among CiM designs considering iso-area constraints, CiM primitives with row/column multiplexing were found to be more susceptible to performance drops compared to those with higher compute parallelism. A notable limitation of CiM was observed in scenarios with low weight reuse, such as when the input is a single vector ($M=1$). In addition, our infrastructure is open-sourced \footnote{\href{https://github.com/tanvisharma/WWWtoCiM}{https://github.com/tanvisharma/WWWtoCiM}}, enabling the inclusion of additional CiM primitives, cost models, and architectures to quickly assess their overall performance and energy efficiency.

 


Overall, our study paves the way for designing CiM accelerators for efficient ML inference. 

\section*{Acknowledgments}
The authors would like to acknowledge North America Qualcomn Innovation Fellowship offered in 2021 for funding the project and inputs from Ramesh Chauhan in the initial stage of the project. Part of the research was also funded by CoCoSys, one of the 7 JUMP centers funded by DARPA and SRC. The authors would also like to thank Aayush Ankit for the brainstorming and discussion sessions.


\appendices

\begin{table}[ht]

\caption{Machine Learning Workload Characteristics}
\begin{tabular}{@{}cccccc@{}}
\textbf{MLWorkload} & \textbf{M} & \textbf{N} & \textbf{K} & \textbf{\#MACs} & \textbf{Algorithm Reuse} \\ 
BERT-Large & 512 & 1024 & 1024 & 536870912 & 512 \\
BERT-Large & 512 & 512 & 1024 & 268435456 & 409.6 \\
BERT-Large & 512 & 1024 & 512 & 268435456 & 409.6 \\
BERT-Large & 512 & 4096 & 1024 & 2147483648 & 630.154 \\
BERT-Large & 512 & 1024 & 4096 & 2147483648 & 630.154 \\
GPT-J & 1 & 4096 & 4096 & 16777216 & 1.999 \\
GPT-J & 2048 & 4096 & 4096 & 34359738368 & 2048 \\
GPT-J & 1 & 2048 & 4096 & 8388608 & 1.999 \\
GPT-J & 1 & 4096 & 2048 & 8388608 & 1.999 \\
GPT-J & 1 & 16384 & 4096 & 67108864 & 1.999 \\
DLRM & 1 & 256 & 512 & 131072 & 1.988 \\
DLRM & 1 & 64 & 256 & 16384 & 1.962 \\
ResNet50 & 12544 & 64 & 147 & 118013952 & 88.860 \\
ResNet50 & 3136 & 64 & 64 & 12845056 & 63.354 \\
ResNet50 & 3136 & 64 & 576 & 115605504 & 113.122 \\
ResNet50 & 3136 & 256 & 64 & 51380224 & 100.755 \\
ResNet50 & 3136 & 64 & 256 & 51380224 & 100.755 \\
ResNet50 & 3136 & 64 & 576 & 115605504 & 113.122 \\
ResNet50 & 3136 & 256 & 64 & 51380224 & 100.755 \\
ResNet50 & 3136 & 64 & 256 & 51380224 & 100.755 \\
ResNet50 & 3136 & 64 & 576 & 115605504 & 113.122 \\
ResNet50 & 3136 & 256 & 64 & 51380224 & 100.755 \\
ResNet50 & 3136 & 128 & 256 & 102760448 & 166.146 \\
ResNet50 & 784 & 128 & 1152 & 115605504 & 200.883 \\
ResNet50 & 784 & 512 & 128 & 51380224 & 181.141 \\
ResNet50 & 784 & 128 & 512 & 51380224 & 181.141 \\
ResNet50 & 784 & 128 & 1152 & 115605504 & 200.883 \\
ResNet50 & 784 & 512 & 128 & 51380224 & 181.141 \\
ResNet50 & 784 & 128 & 512 & 51380224 & 181.141 \\
ResNet50 & 784 & 128 & 1152 & 115605504 & 200.883 \\
ResNet50 & 784 & 512 & 128 & 51380224 & 181.141 \\
ResNet50 & 784 & 128 & 512 & 51380224 & 181.141 \\
ResNet50 & 784 & 128 & 1152 & 115605504 & 200.883 \\
ResNet50 & 784 & 512 & 128 & 51380224 & 181.141 \\
ResNet50 & 784 & 256 & 512 & 102760448 & 280.313 \\
ResNet50 & 196 & 256 & 2304 & 115605504 & 211.812 \\
ResNet50 & 196 & 1024 & 256 & 51380224 & 200.303 \\
ResNet50 & 196 & 256 & 1024 & 51380224 & 200.303 \\
ResNet50 & 196 & 256 & 2304 & 115605504 & 211.812 \\
ResNet50 & 196 & 1024 & 256 & 51380224 & 200.303 \\
ResNet50 & 196 & 256 & 1024 & 51380224 & 200.303 \\
ResNet50 & 196 & 256 & 2304 & 115605504 & 211.812 \\
ResNet50 & 196 & 1024 & 256 & 51380224 & 200.303 \\
ResNet50 & 196 & 256 & 1024 & 51380224 & 200.303 \\
ResNet50 & 196 & 256 & 2304 & 115605504 & 211.812 \\
ResNet50 & 196 & 1024 & 256 & 51380224 & 200.303 \\
ResNet50 & 196 & 256 & 1024 & 51380224 & 200.303 \\
ResNet50 & 196 & 256 & 2304 & 115605504 & 211.812 \\
ResNet50 & 196 & 1024 & 256 & 51380224 & 200.303 \\
ResNet50 & 196 & 256 & 1024 & 51380224 & 200.303 \\
ResNet50 & 196 & 256 & 2304 & 115605504 & 211.812 \\
ResNet50 & 196 & 1024 & 256 & 51380224 & 200.303 \\
ResNet50 & 196 & 512 & 1024 & 102760448 & 249.012 \\
ResNet50 & 49 & 512 & 4608 & 115605504 & 88.581 \\
ResNet50 & 49 & 2048 & 512 & 51380224 & 87.529 \\
ResNet50 & 49 & 512 & 2048 & 51380224 & 87.529 \\
ResNet50 & 49 & 512 & 4608 & 115605504 & 88.581 \\
ResNet50 & 49 & 2048 & 512 & 51380224 & 87.529 \\
ResNet50 & 49 & 512 & 2048 & 51380224 & 87.529 \\
ResNet50 & 49 & 512 & 4608 & 115605504 & 88.581 \\
ResNet50 & 49 & 2048 & 512 & 51380224 & 87.529 \\
ResNet50 & 1 & 1000 & 2048 & 2048000 & 1.997 \\
\end{tabular}

\end{table}
\section{Extra Results}
The main manuscript showcases general matrix multiplication (GEMM) analysis only for Digital6T primitive because of its superior compute parallelism compared to other CiM primitives under iso-area constraints. For the sake of completeness, this section expands on the energy-efficiency and performance results including all CiM primitives for regular (square shaped) GEMMS.

\begin{figure*}[t]
    \centering
    \includegraphics[width=\textwidth]{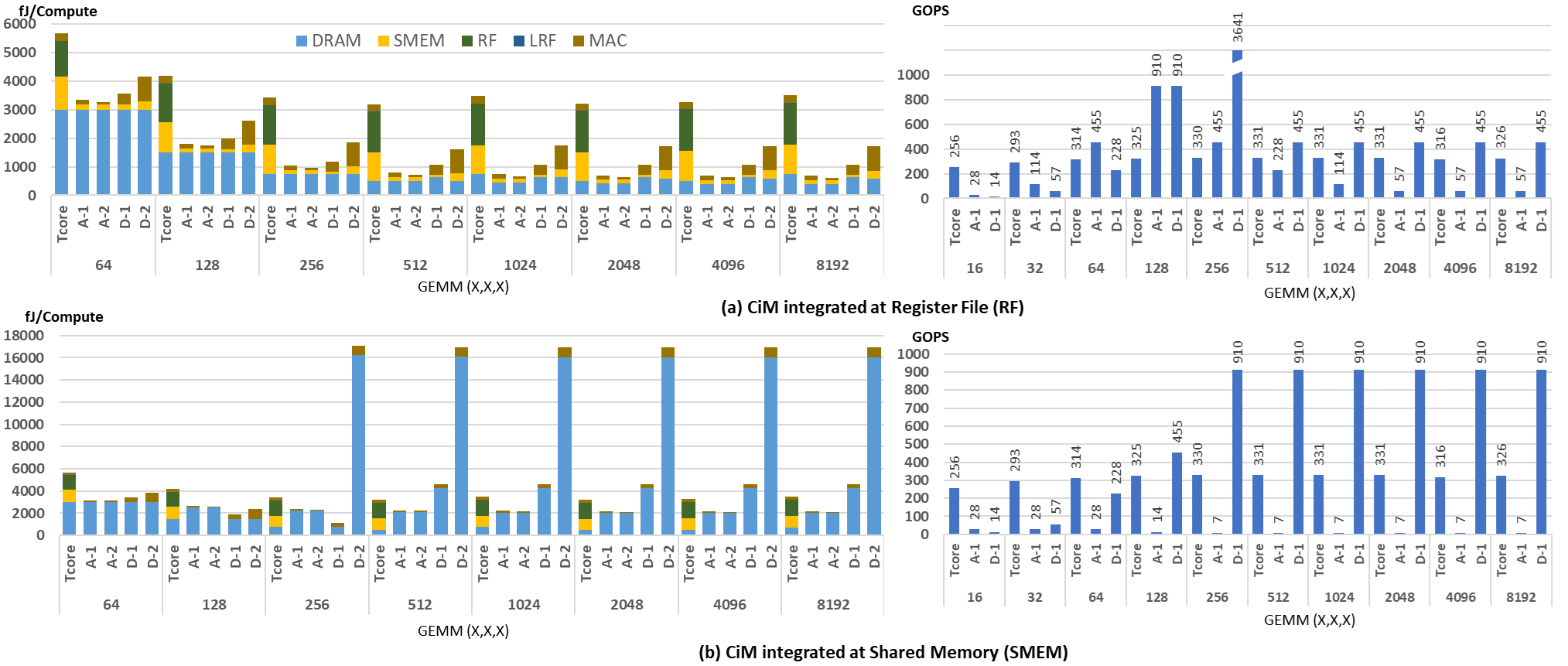}
    \caption{Energy-efficiency ($fJ/compute$) and throughput ($GOPS$) for different GEMM shapes varying from $(64,64,64)$ to $(8192, 8192, 8192)$ with baseline tensor core architecture (Tcore) and CiM configurations - Analog6T (A-1), Analog8T (A-2), Digital6T (D-1) and Digital8T (D-2) under iso-area constraints. The throughput results of A-2 and D-2 are not included in the graph because of their extremely low performance, resulting from lower compute parallelism and high compute latency (Table IV in original manuscript).}
    \label{fig:results-energy}

        \vspace{-4mm}
\end{figure*}
The energy efficiency and observed performance of different CiM primitives when integrated at register file level is shown in Fig. \ref{fig:results-energy}(a). The energy efficiency graph on the left illustrates the distribution of energy across memory levels and compute (MAC) for the different architectures. As the GEMM size increases from $64$ to $8192$, energy per compute reduces for all architectures because the main memory (DRAM) access cost gets amortized over more number of compute operations (blue sub-bars). Moreover, the energy consumption of CiM primitives approximately plateaus after $GEMM (512, 512, 512)$ from optimum utilization of all CiM units. Comparing across different architectures for large GEMM shapes, Analog8T (A-2) beats all other primitives and the baseline tensorcore (Tcore) achieving $\approx620fJ/compute$ while closely competing with Analog6T (A-1) with $\approx700fJ/compute$. Tensor core always consumes higher energy compared to all CiM configurations due to the higher SMEM and RF access cost that gets saved in CiM by performing in-situ computations. CiM integration at RF not only reduces the MAC energy, but also lower level memory accesses by reusing the data stored in RF for compute. The right hand side of Fig. \ref{fig:results-energy}(a) compares the observed throughput across different CiM integrated architectures at register file level. The throughput from both Analog6T (A-1) and Digital6T (D-1) increases with the GEMM size until they reach a sweet point at $X=128$ and $X=256$ respectively. This is in accordance with the results shown in the main manuscript which concludes that the peak throughput is dependent on the number of parallel CiM units and the shape of CiM primitive. For further increase in the GEMM size, throughput from A-1 keeps decreasing because further computations are run in a sequential fashion in the same CiM primitive (an artifact of Rh and Ch). On the other hand, D-1 can fit not only more CiM primitives under iso-area, but offers higher compute parallelism. Therefore, D-1 performance saturates at $455GFLOPS$ while A-1 performance saturates at $57GLOPS$. 

For CiM integration at the shared memory level, the left hand side of Fig. \ref{fig:results-energy}(b) shows the CiM integration can result in higher energy consumption than the baseline tensor core architecture depending on the GEMM size. For instance, the compute energy for D-2 CiM explodes to $\approx4.4\times$ of tensor core energy consumption once the mapping spills the dimensions to DRAM, increasing the cost of partial sum reduction and weight reloading. The varying benefits of CiM integrated at SMEM also brings up the role of CiM primitive shape and parallelism. As an example, GEMM shape with dimension $256$ consumes lowest energy with D-1 CiM configuration due to its shape, indicating that best energy-efficiency CiM (A-2 in our experiments) may not always achieve the best system level energy efficiency. However, the energy consumption of D-1 increase after $256$ due to increase in DRAM access cost from smaller column capacity of D-1 ($16$) compared to A-1 or A-2 ($64$). Looking at the right hand side of the graph, we observe the same trend of throughput hitting maximum value for each architecture for large GEMM shapes. For CiM at SMEM, D-1 exceeds the tensorcore performance after $X=128$ reaching upto $3\times$ higher throughput by storing and reducing larger matrices in memory. Interestingly, A-1 does not perform well in SMEM configuration due to limited reuse opportunities with current mapping scheme. Since Analog-1 has lower parallelism compared to D-1, it results in accessing DRAM more for partial sum reductions, resulting in high bandwidth throttling.

\section{GEMM Shapes}
The exact shapes of the GEMMs found in machine learning workloads during inference with single batch size are listed in this section. We have listed all the $50$ layers of ResNet, however, some of them are not unique and occur more tha once. Note that, for the given SMEM bandwidth and DRAM bandwidth of 42B/cycle and 32B/cycle, the ridge points in the roofline analysis \cite{williams2009roofline} of Digital6T CiM primitive will be $32.5$ and $42.6$ for SMEM and DRAM, respectively. This is calculated based on the peak throughput coming from $3$ instances of Digital6T CiM primitive, that can fit under iso-area constraints at register file level. Here, ridge point defines the minimum value of algorithmic reuse where the workload nature changes from memory-intensive to compute-intensive in nature. Peak performance for a CiM primitive can be calculated as $2*(R_p * C_p * \#CiM~arrays)/(CiM~latency)~GOPS$. The algorithmic reuse along with bandwidth information can help visualize if workloads will experience memory bandwidth throttling in an ideal scenario.


 

\bibliography{refs}

\begin{thebibliography}{10}

\bibitem{welser2018future}
Jeffrey Welser, Jed~W Pitera, and Cindy Goldberg.
\newblock Future computing hardware for ai.
\newblock In {\em 2018 IEEE International Electron Devices Meeting (IEDM)}, pages 1--3. IEEE, 2018.

\bibitem{kim2023full}
Sehoon Kim, Coleman Hooper, Thanakul Wattanawong, Minwoo Kang, Ruohan Yan, Hasan Genc, Grace Dinh, Qijing Huang, Kurt Keutzer, Michael~W Mahoney, et~al.
\newblock Full stack optimization of transformer inference: a survey.
\newblock {\em arXiv preprint arXiv:2302.14017}, 2023.

\bibitem{wulf1995hitting}
Wm~A Wulf and Sally~A McKee.
\newblock Hitting the memory wall: Implications of the obvious.
\newblock {\em ACM SIGARCH computer architecture news}, 23(1):20--24, 1995.

\bibitem{verma2019memory}
Naveen Verma, Hongyang Jia, Hossein Valavi, Yinqi Tang, Murat Ozatay, Lung-Yen Chen, Bonan Zhang, and Peter Deaville.
\newblock In-memory computing: Advances and prospects.
\newblock {\em IEEE Solid-State Circuits Magazine}, 11(3):43--55, 2019.

\bibitem{seo2023advances}
Jae-Sun Seo.
\newblock Advances and trends on on-chip compute-in-memory macros and accelerators.
\newblock In {\em 2023 60th ACM/IEEE Design Automation Conference (DAC)}, pages 1--6. IEEE, 2023.

\bibitem{mutlu2022modern}
Onur Mutlu, Saugata Ghose, Juan G{\'o}mez-Luna, and Rachata Ausavarungnirun.
\newblock A modern primer on processing in memory.
\newblock In {\em Emerging Computing: From Devices to Systems: Looking Beyond Moore and Von Neumann}, pages 171--243. Springer, 2022.

\bibitem{9435945}
Myeonggu Kang, Hyeonuk Kim, Hyein Shin, Jaehyeong Sim, Kyeonghan Kim, and Lee-Sup Kim.
\newblock S-flash: A nand flash-based deep neural network accelerator exploiting bit-level sparsity.
\newblock {\em IEEE Transactions on Computers}, 71(6):1291--1304, 2022.

\bibitem{fujiki2019duality}
Daichi Fujiki, Scott Mahlke, and Reetuparna Das.
\newblock Duality cache for data parallel acceleration.
\newblock In {\em Proceedings of the 46th International Symposium on Computer Architecture}, pages 397--410, 2019.

\bibitem{fujiki2022multi}
Daichi Fujiki, Alireza Khadem, Scott Mahlke, and Reetuparna Das.
\newblock Multi-layer in-memory processing.
\newblock In {\em 2022 55th IEEE/ACM International Symposium on Microarchitecture (MICRO)}, pages 920--936. IEEE, 2022.

\bibitem{lockerman2020livia}
Elliot Lockerman, Axel Feldmann, Mohammad Bakhshalipour, Alexandru Stanescu, Shashwat Gupta, Daniel Sanchez, and Nathan Beckmann.
\newblock Livia: Data-centric computing throughout the memory hierarchy.
\newblock In {\em ASPLOS}, pages 417--433, 2020.

\bibitem{ampere}
NVIDIA.
\newblock {\em Ampere Architecture}.
\newblock \url{https://developer.nvidia.com/blog/nvidia-ampere-architecture-in-depth}.

\bibitem{buluc2010linear}
Aydin Buluc and John~R Gilbert.
\newblock {\em Linear algebraic primitives for parallel computing on large graphs}.
\newblock Citeseer, 2010.

\bibitem{computesram}
Jingcheng Wang, Xiaowei Wang, Charles Eckert, Arun Subramaniyan, Reetuparna Das, David Blaauw, and Dennis Sylvester.
\newblock A 28-nm compute sram with bit-serial logic/arithmetic operations for programmable in-memory vector computing.
\newblock {\em IEEE Journal of Solid-State Circuits}, 55(1):76--86, 2020.

\bibitem{analog1cim}
Xin Si and et~al.
\newblock A local computing cell and 6t sram-based computing-in-memory macro with 8-b mac operation for edge ai chips.
\newblock {\em IEEE Journal of Solid-State Circuits}, 56(9):2817--2831, 2021.

\bibitem{ali2023cicc}
Mustafa Ali, Indranil Chakraborty, Sakshi Choudhary, Muya Chang, Dong~Eun Kim, Arijit Raychowdhury, and Kaushik Roy.
\newblock A 65 nm 1.4-6.7 tops/w adaptive-snr sparsity-aware cim core with load balancing support for dl workloads.
\newblock In {\em 2023 IEEE Custom Integrated Circuits Conference (CICC)}, pages 1--2, 2023.

\bibitem{digitaltsmc}
Yu-Der Chih et~al.
\newblock 16.4 an 89tops/w and 16.3tops/mm2 all-digital sram-based full-precision compute-in memory macro in 22nm for machine-learning edge applications.
\newblock In {\em 2021 IEEE International Solid- State Circuits Conference (ISSCC)}, volume~64, pages 252--254, 2021.

\bibitem{tmsc2023digital}
Haruki Mori et~al.
\newblock A 4nm 6163-tops/w/b $\mathbf{4790-TOPS/mm^{2}/b}$ sram based digital-computing-in-memory macro supporting bit-width flexibility and simultaneous mac and weight update.
\newblock In {\em 2023 IEEE International Solid- State Circuits Conference (ISSCC)}, pages 132--134, 2023.

\bibitem{isscc2022mfchang}
Ping-Chun Wu et~al.
\newblock A 28nm 1mb time-domain computing-in-memory 6t-sram macro with a 6.6ns latency, 1241gops and 37.01tops/w for 8b-mac operations for edge-ai devices.
\newblock In {\em 2022 IEEE International Solid- State Circuits Conference (ISSCC)}, volume~65, pages 1--3, 2022.

\bibitem{tsmc2022isscc}
Hidehiro Fujiwara et~al.
\newblock A 5-nm 254-tops/w 221-tops/mm2 fully-digital computing-in-memory macro supporting wide-range dynamic-voltage-frequency scaling and simultaneous mac and write operations.
\newblock In {\em 2022 IEEE International Solid- State Circuits Conference (ISSCC)}, volume~65, pages 1--3, 2022.

\bibitem{chetlur2014cudnn}
Sharan Chetlur, Cliff Woolley, Philippe Vandermersch, Jonathan Cohen, John Tran, Bryan Catanzaro, and Evan Shelhamer.
\newblock cudnn: Efficient primitives for deep learning.
\newblock {\em arXiv preprint arXiv:1410.0759}, 2014.

\bibitem{gemmnv}
Nvidia Docs.
\newblock {\em Matrix Multiplication Background User's Guide}.
\newblock https://docs.nvidia.com/deeplearning/performance/dl-performance-matrix-multiplication/index.html, 2020-23.

\bibitem{devic2022pim}
Alexandar Devic, Siddhartha~Balakrishna Rai, Anand Sivasubramaniam, Ameen Akel, Sean Eilert, and Justin Eno.
\newblock To pim or not for emerging general purpose processing in ddr memory systems.
\newblock In {\em Proceedings of the 49th Annual International Symposium on Computer Architecture}, pages 231--244, 2022.

\bibitem{houshmand2023benchmarking}
Pouya Houshmand, Jiacong Sun, and Marian Verhelst.
\newblock Benchmarking and modeling of analog and digital sram in-memory computing architectures.
\newblock {\em arXiv preprint arXiv:2305.18335}, 2023.

\bibitem{naumov2019deep}
Maxim Naumov, Dheevatsa Mudigere, Hao-Jun~Michael Shi, Jianyu Huang, Narayanan Sundaraman, Jongsoo Park, Xiaodong Wang, Udit Gupta, Carole-Jean Wu, Alisson~G Azzolini, et~al.
\newblock Deep learning recommendation model for personalization and recommendation systems.
\newblock {\em arXiv preprint arXiv:1906.00091}, 2019.

\bibitem{parashar2017scnn}
Angshuman Parashar, Minsoo Rhu, Anurag Mukkara, Antonio Puglielli, Rangharajan Venkatesan, Brucek Khailany, Joel Emer, Stephen~W Keckler, and William~J Dally.
\newblock Scnn: An accelerator for compressed-sparse convolutional neural networks.
\newblock {\em ACM SIGARCH computer architecture news}, 45(2):27--40, 2017.

\bibitem{parashar2019timeloop}
Angshuman Parashar, Priyanka Raina, Yakun~Sophia Shao, Yu-Hsin Chen, Victor~A Ying, Anurag Mukkara, Rangharajan Venkatesan, Brucek Khailany, Stephen~W Keckler, and Joel Emer.
\newblock Timeloop: A systematic approach to dnn accelerator evaluation.
\newblock In {\em 2019 IEEE international symposium on performance analysis of systems and software (ISPASS)}, pages 304--315. IEEE, 2019.

\bibitem{kwon2020maestro}
Hyoukjun Kwon, Prasanth Chatarasi, Vivek Sarkar, Tushar Krishna, Michael Pellauer, and Angshuman Parashar.
\newblock Maestro: A data-centric approach to understand reuse, performance, and hardware cost of dnn mappings.
\newblock {\em IEEE micro}, 40(3):20--29, 2020.

\bibitem{mei2020zigzag}
Linyan Mei, Pouya Houshmand, Vikram Jain, Sebastian Giraldo, and Marian Verhelst.
\newblock Zigzag: A memory-centric rapid dnn accelerator design space exploration framework.
\newblock {\em arXiv preprint arXiv:2007.11360}, 2020.

\bibitem{horowitz20141}
Mark Horowitz.
\newblock 1.1 computing's energy problem (and what we can do about it).
\newblock In {\em 2014 IEEE international solid-state circuits conference digest of technical papers (ISSCC)}, pages 10--14. IEEE, 2014.

\bibitem{shanbhag2022benchmarking}
Naresh~R Shanbhag and Saion~K Roy.
\newblock Benchmarking in-memory computing architectures.
\newblock {\em IEEE Open Journal of the Solid-State Circuits Society}, 2:288--300, 2022.

\bibitem{impulse}
Amogh Agrawal, Mustafa Ali, Minsuk Koo, Nitin Rathi, Akhilesh Jaiswal, and Kaushik Roy.
\newblock Impulse: A 65-nm digital compute-in-memory macro with fused weights and membrane potential for spike-based sequential learning tasks.
\newblock {\em IEEE Solid-State Circuits Letters}, 4:137--140, 2021.

\bibitem{ankit2019puma}
Aayush Ankit, Izzat~El Hajj, Sai~Rahul Chalamalasetti, Geoffrey Ndu, Martin Foltin, R~Stanley Williams, Paolo Faraboschi, Wen-mei~W Hwu, John~Paul Strachan, and Kaushik Roy.
\newblock Puma: A programmable ultra-efficient memristor-based accelerator for machine learning inference.
\newblock In {\em ASPLOS}, 2019.

\bibitem{tsmc7nmanalog}
Qing Dong, Mahmut~E. Sinangil, Burak Erbagci, Dar Sun, Win-San Khwa, Hung-Jen Liao, Yih Wang, and Jonathan Chang.
\newblock 15.3 a 351tops/w and 372.4gops compute-in-memory sram macro in 7nm finfet cmos for machine-learning applications.
\newblock In {\em 2020 IEEE International Solid-State Circuits Conference - (ISSCC)}, pages 242--244, 2020.

\bibitem{ali2020imac}
Mustafa Ali, Akhilesh Jaiswal, Sangamesh Kodge, Amogh Agrawal, Indranil Chakraborty, and Kaushik Roy.
\newblock Imac: In-memory multi-bit multiplication and accumulation in 6t sram array.
\newblock {\em IEEE Transactions on Circuits and Systems I: Regular Papers}, 67(8):2521--2531, 2020.

\bibitem{stillmaker2017scaling}
Aaron Stillmaker and Bevan Baas.
\newblock Scaling equations for the accurate prediction of cmos device performance from 180 nm to 7 nm.
\newblock {\em Integration}, 58:74--81, 2017.

\bibitem{nagel2019data}
Markus Nagel, Mart~van Baalen, Tijmen Blankevoort, and Max Welling.
\newblock Data-free quantization through weight equalization and bias correction.
\newblock In {\em Proceedings of the IEEE/CVF International Conference on Computer Vision}, pages 1325--1334, 2019.

\bibitem{dettmers2022llm}
Tim Dettmers, Mike Lewis, Younes Belkada, and Luke Zettlemoyer.
\newblock Llm. int8 (): 8-bit matrix multiplication for transformers at scale.
\newblock {\em arXiv preprint arXiv:2208.07339}, 2022.

\bibitem{iccad_2019_accelergy}
Yannan~N. Wu, Joel~S. Emer, and Vivienne Sze.
\newblock {Accelergy: An Architecture-Level Energy Estimation Methodology for Accelerator Designs}.
\newblock In {\em {IEEE/ACM International Conference On Computer Aided Design (ICCAD)}}, {2019}.

\bibitem{he2015deep}
Kaiming He, Xiangyu Zhang, Shaoqing Ren, and Jian Sun.
\newblock Deep residual learning for image recognition. corr abs/1512.03385 (2015), 2015.

\bibitem{deng2009imagenet}
Jia Deng, Wei Dong, Richard Socher, Li-Jia Li, Kai Li, and Li~Fei-Fei.
\newblock Imagenet: A large-scale hierarchical image database.
\newblock In {\em 2009 IEEE conference on computer vision and pattern recognition}, pages 248--255. Ieee, 2009.

\bibitem{devlin2018bert}
Jacob Devlin, Ming-Wei Chang, Kenton Lee, and Kristina Toutanova.
\newblock Bert: Pre-training of deep bidirectional transformers for language understanding.
\newblock {\em arXiv preprint arXiv:1810.04805}, 2018.

\bibitem{gpt-j}
Ben Wang and Aran Komatsuzaki.
\newblock {GPT-J-6B: A 6 Billion Parameter Autoregressive Language Model}.
\newblock \url{https://github.com/kingoflolz/mesh-transformer-jax}, May 2021.

\bibitem{saxena2022towards}
Utkarsh Saxena, Indranil Chakraborty, and Kaushik Roy.
\newblock Towards adc-less compute-in-memory accelerators for energy efficient deep learning.
\newblock In {\em 2022 Design, Automation \& Test in Europe Conference \& Exhibition (DATE)}, pages 624--627. IEEE, 2022.

\bibitem{wu202322nm}
Ping-Chun Wu, Jian-Wei Su, Li-Yang Hong, Jin-Sheng Ren, Chih-Han Chien, Ho-Yu Chen, Chao-En Ke, Hsu-Ming Hsiao, Sih-Han Li, Shyh-Shyuan Sheu, et~al.
\newblock A 22nm 832kb hybrid-domain floating-point sram in-memory-compute macro with 16.2-70.2 tflops/w for high-accuracy ai-edge devices.
\newblock In {\em 2023 IEEE International Solid-State Circuits Conference (ISSCC)}, pages 126--128. IEEE, 2023.

\bibitem{yue202328nm}
Jinshan Yue, Chaojie He, Zi~Wang, Zhaori Cong, Yifan He, Mufeng Zhou, Wenyu Sun, Xueqing Li, Chunmeng Dou, Feng Zhang, et~al.
\newblock A 28nm 16.9-300tops/w computing-in-memory processor supporting floating-point nn inference/training with intensive-cim sparse-digital architecture.
\newblock In {\em 2023 IEEE International Solid-State Circuits Conference (ISSCC)}, pages 1--3. IEEE, 2023.

\bibitem{williams2009roofline}
Samuel Williams, Andrew Waterman, and David Patterson.
\newblock Roofline: an insightful visual performance model for multicore architectures.
\newblock {\em Communications of the ACM}, 52(4):65--76, 2009.

\end{thebibliography}












\section{Biography Section}
\vspace{-40pt}
\begin{IEEEbiography}[{\includegraphics[width=1in,height=1.25in,clip,keepaspectratio]{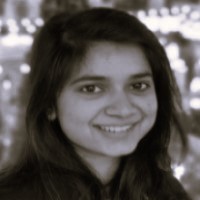}}]{Tanvi Sharma} received her bachelor’s degree from Indian Institute of Technology, Roorkee in 2018 and worked as a Digital Design Engineer at Texas Instruments before joining Purdue University in 2019. She is in the direct PhD program at Purdue, under the guidance of Professor Kaushik Roy. She has been a recipient of Qualcomn Innovation Fellowship in 2021, and MLSys Rising Stars Award in 2023. Her research interests lie at the intersection of machine learning and systems, with focus on developing energy efficient ML hardware accelerators using compute-in-memory solutions.
\end{IEEEbiography}

\vspace{2pt}

\begin{IEEEbiography}[{\includegraphics[width=1in,height=1.25in,clip,keepaspectratio]{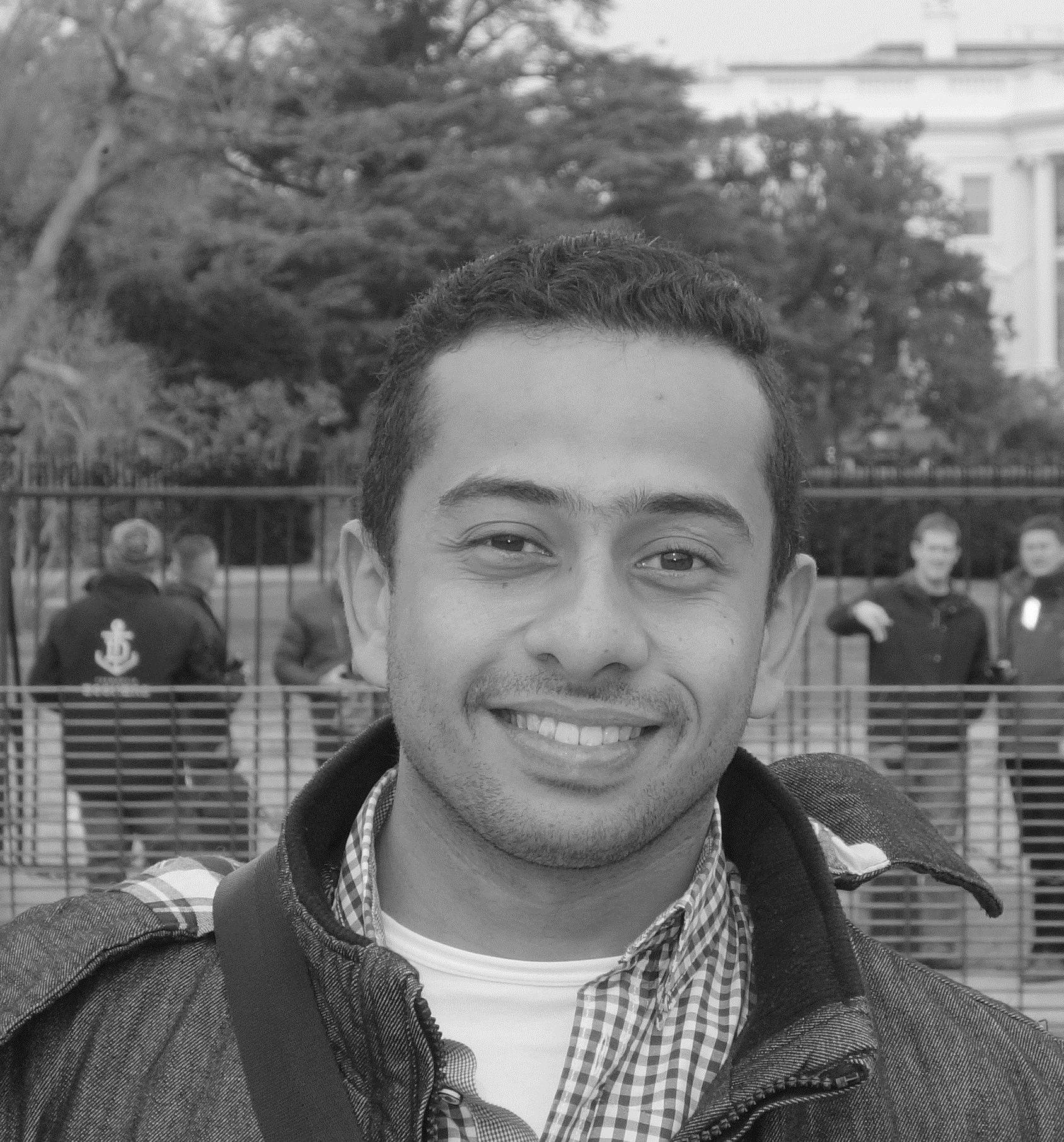}}]{Mustafa Ali} received his B.Sc. ans M.Sc. degrees in Electrical Engineering from MTC, Cairo, Egypt in 2011, 2016 respectively.
He worked on flexible electronics applications using TFTs in his M.Sc from 2014 to 2016. Additionally, he worked as a TA and RA at MTC from 2013 to 2017.
Mustafa was also a hardware and embedded systems engineer at Integreight, Inc. from 2012 to 2017. He joined the Nano-electronics Research Lab (NRL), Purdue University in Spring 2018 to Fall 2022 pursuing his Ph.D. under the guidance of Prof. Roy. His research interest lied in innovative HW accelerating of ML workloads inculding compute in memory. He is currently a hardware engineer at Microsoft Azure Hardware Systems and Infrastructure.
\end{IEEEbiography}

\vspace{4pt}

\begin{IEEEbiography}[{\includegraphics[width=1in,height=1.25in,clip,keepaspectratio]{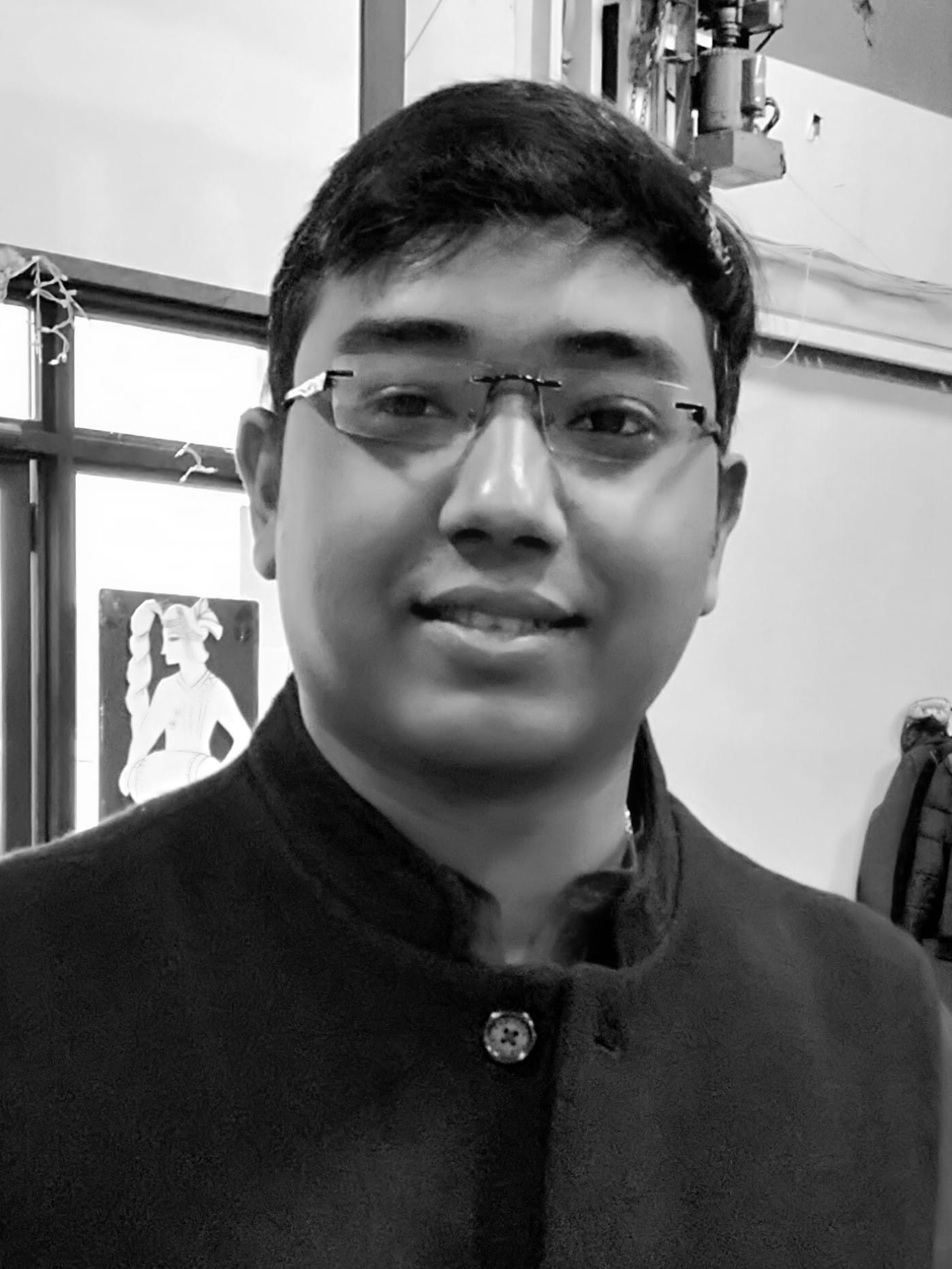}}]{Indranil Chakraborty} is currently a Hardware Engineer at Google, Sunnyvale, California. He received the B.Engg. degree in Electronics and Telecommunication Engineering from Jadavpur University, Kolkata, India, in 2013, an M.Tech. degree in Electrical Engineering from Indian Institute of Technology Bombay, Mumbai, India, in 2016 and Ph.D. degree in 2021 at Nanoelectronics Research Laboratory, Purdue University, West Lafayette, IN, USA. His primary research interests lie in architecture and design of hardware accelerators for machine-learning workloads using CMOS and emerging technologies.
\end{IEEEbiography}

\vspace{4pt}

\begin{IEEEbiography}[{\includegraphics[width=1in,height=1.25in,clip,keepaspectratio]{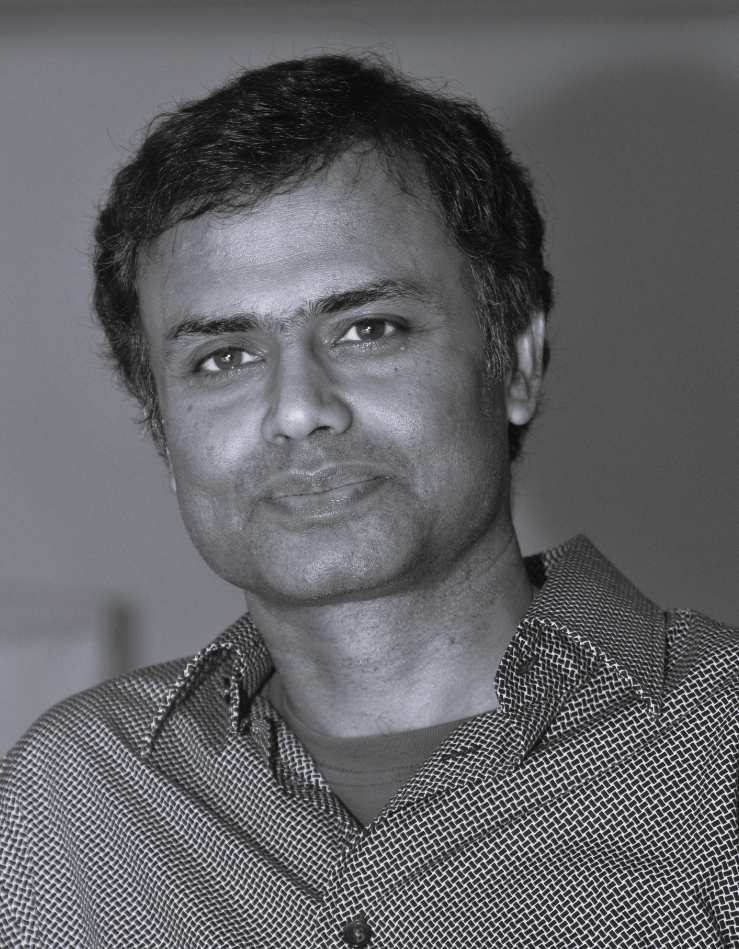}}]{Kaushik Roy} is the Edward G. Tiedemann, Jr., Distinguished Professor of Electrical and Computer Engineering at Purdue University. He received his BTech from Indian Institute of Technology, Kharagpur, PhD from University of Illinois at Urbana-Champaign in 1990 and joined the Semiconductor Process and Design Center of Texas Instruments, Dallas, where he worked for three years on FPGA architecture development and low-power circuit design. His current research focuses on cognitive algorithms, circuits and architecture for energy-efficient neuromorphic computing/ machine learning, and neuro-mimetic devices. Kaushik has supervised more than 100 PhD dissertations and his students are well placed in universities and industry. He is the co-author of two books on Low Power CMOS VLSI Design (John Wiley \& McGraw Hill). 
\end{IEEEbiography}


\vfill

\end{document}